\documentstyle[12pt]{article}
\SetSymbolFont{largesymbols}{normal}{OMX}{lcmex}{m}{n}
  
\catcode`@=11
\def\secteqno{\@addtoreset{equation}{section}%
\def\theequation{\thesection.\arabic{equation}}}
\catcode`@=12
\secteqno
\def\dd{\hbox{\,\Large$\triangleright$}}
\newcommand{\be}{\begin{equation}}
\newcommand{\ee}{\end{equation}}
\newcommand{\bea}{\begin{eqnarray}}
\newcommand{\eea}{\end{eqnarray}}
\newcommand{\bref}[1]{(\ref{#1})}
\newcommand{\nn}{\nonumber}

\newcommand{\slP}{/ {\hskip-0.27cm{P}}}
\newcommand{\slSigma}{/ {\hskip-0.27cm{\Sigma}}}
\newcommand{\slS}{/ {\hskip-0.27cm{S}}}

\begin{document}

\begin{flushright}
\parbox{4.2cm}
{2014,~March 16 \\
KEK-TH-1713 \hfill \\YITP-SB-14-7
\hfill \\
}
\end{flushright}

\vspace*{1.1cm}

\begin{center}
 {\Large\bf   Superspace with manifest T-duality}\\
 {\Large\bf  from type II superstring}
\end{center}
\vspace*{1.5cm}
\centerline{\large Machiko Hatsuda$^{\dagger^\natural }$\footnote{mhatsuda@post.kek.jp
}, Kiyoshi Kamimura$^{\ast }$\footnote{kamimura@ph.sci.toho-u.ac.jp}
and Warren Siegel$^\star$
\footnote{siegel@insti.physics.sunysb.edu
,
http://insti.physics.sunysb.edu/{\tt \~{}}siegel/plan.html}
}
\begin{center}
$^{\dagger}$\emph{Physics Division, Faculty of Medicine,
 Juntendo University, Chiba 270-1695, Japan}
\\
$^{\natural}$\emph{KEK Theory Center, High Energy Accelerator Research 
Organization,\\
Tsukuba, Ibaraki 305-0801, Japan} 
\\
$^{\ast}$\emph{Department of Physics, Toho University, Funabashi 274-8510, Japan
}\\
$^\star$\emph{C. N. Yang Institute for Theoretical Physics
State University of New York, Stony Brook, NY 11794-3840}
\vspace*{0.5cm}
\\

\end{center}

\vspace*{1cm}

\centerline{\bf Abstract}
A superspace formulation  of type II superstring background
with manifest  T-duality symmetry  
is presented.
This manifestly T-dual formulation is constructed in a space spanned by 
 two sets of nondegenerate super-Poincar\'{e} algebras.
Supertorsion constraints are obtained from consistency of
the $\kappa$-symmetric Virasoro constraints.
All superconnections and vielbein fields are solved in terms of
a prepotential which is one of the vielbein components.
 AdS$_5\times$S$^5$ background is explained 
in this formulation.

\vspace*{0.5cm}

 \vfill 

\thispagestyle{empty}
\setcounter{page}{0}
\newpage

\section{Introduction}

T-duality symmetry, which is the symmetry of
 the low energy effective theory of string, 
was manifestly realized by spacetime coordinates doubled 
firstly in \cite{Siegel:1993xq}.
T-duality has brought a variety of studies on 
the generalized complex geometry 
 triggered by   \cite{Hitchin:2004ut}.
Both aspects of T-duality are characterized by
generalized diffeomorphism with manifest T-duality.  
Various topics in the early stage  are  for example in
\cite{Hull:2004in,Grana:2005jc}, and 
recent development is in review articles for example 
\cite{Hohm:2013bwa,Berman:2014jba}.
Supersymmetric T-duality theories with Ramond-Ramond fields
 have been also developed 
\cite{Hassan:1999bv,Fukuma:1999jt,Hohm:2011zr,Jeon:2012kd}.

In this paper we propose a superspace formulation 
  of type II superstring backgrounds
with the manifest  T-duality symmetry.
 The space is spanned by two sets of 
 nondegenerate super-Poincar\'{e} algebras.
The nondegeneracy of the group 
 is required by consistency of the affine Lie algebra.
A nondegenerate pair of  supertranslation generators $(D_\mu,~\Omega^\mu)$
  was introduced in   \cite{Siegel:85,Siegel:1985xj},  
  while a nondegenerate pair of Lorentz generators  $(S_{mn},~\Sigma^{mn})$
  was introduced in \cite{Hatsuda:2001xf,Bonanos:2009wy,
  Siegel:2011sy,Polacek:2013nla}.
  Type II supergravity theories in superspace approach
   were presented in 
 \cite{Howe:1981gz,Howe:1983sra,Berkovits:2001ue}
 where different superspaces are used;
$(D,P)$ superspace is used in  \cite{Howe:1981gz,Howe:1983sra},
$(S,D,P,\Omega)$ superspace is used in  \cite{Berkovits:2001ue}.
We use  $(S,D,P,\Omega,\Sigma)$ superspace.
The difference leads to  different sets of auxiliary fields 
for the different symmetries,
despite of the same physical degrees of freedom.

There are gauge symmetries generated by
 the generalized Lie bracket \cite{Siegel:1993xq}.
The  gauge symmetry of the doubled  nondegenerate super-Poincar\'{e} space
includes the general coordinate invariance, the gauge symmetry of NS-NS  $B$ field, 
two local supersymmetries,  and two local Lorentz's.
Curved backgrounds are introduced in a covariant way by 
a vielbein superfield $E_{\underline{A}\underline{M}}$.
Supersymmetric  vielbein components include 
the R-R field strength  $E_{\underline{\Omega\Omega}}$ in addition to 
the gravitational field $E_{(\underline{PP})}$,
the NS-NS $B$ field  $E_{[\underline{PP}]}$ and 
the NS-NS field strength  $E_{\underline{P\Sigma}}$, 
which are mixed under the T-duality transformation.
The T-duality symmetry acts linearly on the vielbein field in a string or a brane  
Hamiltonian  \cite{Hatsuda:2012uk,Hatsuda:2012vm},
 while it acts fractional linearly on the gravitational field $G$
 and the $B$ field, $G+B$, in the low energy effective theory.
Therefore we begin by the superstring mechanics 
to construct a manifestly T-dual formulation of the type II supergravity.

In a  superspace approach supersymmetry is manifest,
 but a consistent set of torsion constraints is nontrivial.
Torsion constraints are derived from the $\kappa$-symmetry of the superstring.
It was shown that the $\kappa$-symmetric Virasoro condition in curved superspace
is equal to the supergravity equation of motion \cite{Witten:1985nt,Shapiro:1986xp}.
We construct $\kappa$-symmetric Virasoro constraints 
in the nondegenerate super-Poincar\'{e} space. 
Torsion constraints are solved dimension by dimension.
The gauges are chosen by solving torsion constraints so that all of vielbein 
are  fixed in terms of a prepotential, 
$E_{\underline{DD}}=B_{\underline{\mu\nu}}$, which is a spinor-spinor component of the
$B$ field with dimension $-1$.

We examine an AdS space case as an curved background example. 
The covariant derivatives of superstring in the AdS$_5\times$S$^5$ background 
were computed in \cite{Hatsuda:2001xf} showing the  $\kappa$-symmetric
Virasoro constraints up to the Lorentz constraint.
In this paper we recompute the algebra with manifest Lorentz current $S$,
and reconstruct $\Omega$ and $\Sigma$ currents 
in such a way that supertorsions are totally graded antisymmetric. 
Then torsion constraints in the manifestly T-dual formulation
are confirmed by comparing with obtained supertorsions and supercurvature tensors.

The procedure of the manifestly T-dual formulation \cite{Polacek:2013nla}
in superspace version is the following:
\begin{enumerate}
  \item {Extend a Lie algebra to an affine Lie algebra.
 
 Begin with a coset G/H where  G must have a nondegenerate group metric.
Double the generators, which become the basis of the T-duality symmetry rotation. 
Generalized diffeomorphism is generated by the zero-mode of the affine Lie algebra.
     }
     \item{Make covariant derivatives with vielbein $E$.
     
Superconnection fields in the covariant derivatives 
are recognized as H components of the vielbein, $E_{\underline{A}}{}^{\rm H}$.     
Impose coset constraints  and 
an orthogonal condition on vielbein field
to reduce vielbein field contents. 
 }
\item{  Constrain torsions from the $\kappa$-symmetry.
  
Construct a set of $\kappa$-symmetric Virasoro constraints,
then examine its consistency.
Supercurvature tensors are recognized as H components of torsion,
$T_{\underline{AB}}{}^{\rm H}$.  
     }
     \item{Break manifest T-duality symmetry to hidden one.
     
     Impose dimensional reduction conditions  using with
      symmetry generators which commute with covariant derivatives.
Gauge fixings allow to obtain the usual set of coordinates.}
\end{enumerate}

The organization of the paper is the following:
In section 2  notation of derivatives and indices is summarized. 
Then the affine super-Poincar\'{e} algebra  
generated by $(S,D,P,\Omega,\Sigma)$
is presented. 
In section 3 the $\kappa$-symmetry constraints as well as Virasoro constraints 
for type II superstrings are presented.
In section 4 the manifestly T-dual formalism of type II superstring background 
 is presented.
The vielbein field, embedded in an element of OSp(d$^2$,d$^2\mid$
2$^{d/2}$, 2$^{d/2}$), includes superconnections 
which are usually treated separately from vielbein.
Torsion constraints and supercurvature tensors are
obtained, where all superconnections 
are solved in terms of a prepotential.
The prepotential superfield is a component of the vielbein with dimension $-1$
which is the lowest dimension of fields.
In section 5 the superstring current algebra  in the
 AdS$_5\times$S$^5$ background is presented.
Supertorsions and supercurvature tensors are determined. 
Dimensional reduction constraints
are imposed as well as  gauge fixing conditions  of 
local Lorentz constraints and the section condition
in order to reduce to the usual space.

 \par
 \vskip 6mm

\section{Type II nondegenerate superspace}
 
 We consider a superspace which is defined by 
 covariant derivatives of 
 two sets of nondegenerate super-Poincar\'{e} algebras.
 At first a notation of several kinds of covariant derivatives 
is listed. 
Then  notation of algebras and indices is listed.   
Next the nondegenerate super-Poincar\'{e} algebras are presented.

\subsection{Notation}

\begin{description}
  \item[Covariant derivatives] 
\end{description}

Several kinds of covariant derivatives are denoted as
\bea
\begin{array}{cccccc}
&{\rm Lie~algebra}&\rightarrow&{\rm particle}&\rightarrow&{\rm string}\\
{\rm Flat}&G_I&
&\mathring{\nabla}_I&&\mathring{\dd}_I\\
\downarrow&& &&&\\
{\rm Curved}&&
&\nabla_A=E_A{}^I\mathring{\nabla}_I&
& \dd_A=E_A{}^I\mathring{\dd}_I
\end{array}
\eea
Hamiltonians for a particle and a string
 in curved background are given by  bilinears of ${\nabla}_A$'s and
 ${\dd}_A$'s respectively, where $E_A{}^I$ is vielbein field. 

\begin{itemize}

  \item {Particle algebra in flat space

  For a Lie algebra $G_I$ 
\bea
&&[G_{I},G_{J}\}=if_{IJ}{}^KG_K ~~~,\nn
\eea  
symmetry generator  $\tilde{\nabla}$
 and covariant derivative  $\mathring{\nabla}$ 
are constructed
by  introducing a group element $g(Z^I)$ 
with particle coordinates $Z^I$. 
The coefficient of Right invariant one form
is denoted by $L_M{}^I$ used in  the left action generator 
and vice versa by $R_M{}^I$ used in the right action generator 
;
\bea
~~~{\renewcommand{\arraystretch}{1.8}
\begin{array}{lll}
&{\rm Symmetry~generator}:&\tilde{\nabla}{}_I=L_I{}^M
\displaystyle\frac{1}{i}\partial_M~,~
(dg)g^{-1}=idZ^ML_M{}^IG_I\\
&{\rm Covariant ~derivative}:&\mathring{\nabla}{}_I
=R_I{}^M\displaystyle\frac{1}{i}\partial_M~,~
g^{-1}(dg)=idZ^MR_M{}^IG_I=J^IG_I
\end{array}}\nn
\eea
\bea
&&{\rm Algebra}\nn\\
&&
[\tilde{\nabla}_{I},\tilde{\nabla}_{J}\}=if_{IJ}{}^K
\tilde{\nabla}_K~,~
~~~[\mathring{\nabla}{}_{I},\mathring{\nabla}{}_{J}\}=-if_{IJ}{}^K
\mathring{\nabla}{}_K~,~
[\mathring{\nabla}{}_{I},\tilde{\nabla}_{J}\}=0~~\nn
\eea 
where $\left[*,*\right\}$ is 
a graded bracket which is anticommuting for two fermions and 
commuting for others. 
}

\item{String algebra in flat space

Affine Lie algebra is obtained by stringy extension as
$Z(\tau,\sigma)$ 
\bea
&{\rm Symmetry~generator}:&\tilde{{\dd}}{}_I=
L_I{}^M\left(\frac{1}{i}\partial_M+\partial_\sigma Z^NB_{NM}\right)
-\partial_\sigma Z^ML_M{}^J\mathring{\eta}_{JI}~~\nn\\
&{\rm Flat~current}:&
\mathring{\dd}{}_I=R_I{}^M
\left(
\frac{1}{i}\partial_M+\partial_\sigma Z^NB_{NM}\right)
+\partial_\sigma Z^M R_M{}^J\mathring{\eta}_{JI}\nn
\eea
\bea
&&{\rm Algebra}\nn\\
&&~~~{\renewcommand{\arraystretch}{1.6}
\begin{array}{ccl}
 \lbrack\tilde{{\dd}}_{I}(1),\tilde{{\dd}}_{J}(2) \}&=&i
 f_{IJ}{}^K\tilde{{\dd}}_K\delta(2-1)
+i\mathring{\eta}_{IJ}\partial_{\sigma}\delta(2-1)\nn\\
 \lbrack \mathring{\dd}{}_{I}(1),\mathring{\dd}{}_{J}(2) \}&=&-if_{IJ}{}^K
\mathring{\dd}{}_K\delta(2-1)-i\mathring{\eta}_{IJ}\partial_{\sigma}\delta(2-1)
\nn\\
\lbrack\mathring{\dd}{}_{I}(1),\tilde{{\dd}}_{J}(2) \}&=&0~~
\end{array}
 }~~~~~~~~~~~~~
\eea 
with $1=\sigma_1$,  $2=\sigma_2$,
 $\delta(2-1)=\delta(\sigma_1-\sigma_2)$ and
 $\partial_\sigma\delta(2-1)=\partial_{\sigma_2}\delta(\sigma_2-\sigma_1)$.
There appear Schwinger terms including $\partial_\sigma\delta$. 
A group metric
$\mathring{\eta}_{IJ}$ is a graded symmetric 
nondegenerate constant matrix. 
 }

\end{itemize}

\begin{itemize}
 \item{Particle algebra in curved space

By introducing background fields through vielbein 
 $E_{A}{}^{J}(Z)$,
 the covariant derivatives in a curved background and the algebra become
 \bea
 ~{\renewcommand{\arraystretch}{1.6}
 \begin{array}{lll}
{\rm  Covariant ~derivative}:
&\nabla_{A}=E_{A}
{}^{J}(Z)\mathring{\nabla}{}_{J}&\nn\\
{\rm Algebra}:&[\nabla_{A},\nabla_{B}\}=-i
T_{AB}{}^{C}\nabla_{C}& ~~~.
\end{array}}\nn
\eea
The torsion $T_{AB}{}^{C}$ is a function of
vielbein fields  whose flat limit is the structure constant
$f_{IJ}{}^K$.
}
\item{String algebra in curved space

The covariant derivatives in a curved background
and the affine algebra become
  \bea
  ~{\renewcommand{\arraystretch}{1.6}
   \begin{array}{lll}
{\rm  Curved~current}:
 &
 {\dd}_{A}=
E_{A}{}^{J}(Z(\tau,\sigma))
 \mathring{\dd}{}_{J}&
 \nn
 \\
 {\rm Algebra} &
  [{\dd}_{A},{\dd}_{B} \}
=-iT_{AB} {}^{C} 
{\dd}_{C}\delta(2-1)
-i\eta_{AB}\partial_\sigma\delta(2-1)& ~~~.\nn
\end{array}}
 \eea
The torsion $T_{AB}{}^C$ and the group metric $\eta_{AB}$ are functions of vielbein fields in general.
We will impose  $\eta_{AB}$ to be constant
$\mathring{\eta}_{IJ}$.
}
\end{itemize}
\begin{description}
  \item[Nondegenerate super-Poincar\'{e}] 
\end{description}

We consider the  super-Poincar\'{e} group with introducing a nondegenerate group metric. 
Closure of the Jacobi identity of an affine algebra requires
a totally  graded  antisymmetric structure constant.
The existence of nondegenerate group  metric allows to totally 
antisymmetric  structure constant, 
$f_{IJK}\equiv f_{IJ}{}^L\eta_{LK}=\frac{1}{3!}f_{[IJK]}$.

 For a superstring in a flat space the NS-NS three form is
 given by a closed three form,
  $H=J^D\wedge J^D \wedge J^P $.
 Contrast to the fact that    $B_{NM}$ field, defined by 
 $\int H=dZ^M dZ^N B_{NM}$, cannot be constant  in conventional superspace,
 the two form potential can be constructed as 
   $B=J^D \wedge J^\Omega$ with $H=dB$ 
   in the nondegenerate superspace.
   
We begin with a space generated by translation generator $p$,
then we add supersymmetry generator $D$
  and Lorentz generator $S$.
They are extended to affine algebras,
and further extended in nondegenerate manner as
\bea
{\renewcommand{\arraystretch}{1.3}
\begin{array}{llll}
&{\rm translation}&\rightarrow~{\rm supertranslation}&\rightarrow~
{\rm super-Poincar\acute{\rm e}}\\
~{\rm particle}&~~p_m&~~~D_\mu,p_m&~~~S_{mn},D_\mu,p_m\\
~\downarrow&&&\\
\begin{array}{l}
{\rm open}\\
{\rm string}
\end{array}&~~P_m=p+\partial_\sigma x
&~~~D_\mu,P_m,\Omega^\mu&~~~S_{mn},D_\mu,P_m\,\Omega^\mu,\Sigma^{mn}\\
~\downarrow&&&\\
\begin{array}{l}
{\rm type~II}\\
{\rm string}
\end{array}&
\left\{\begin{array}{l}
P_m=p+\partial_\sigma x\\
P_{m'}=p-\partial_\sigma x\end{array}\right.
&\left\{\begin{array}{l}
D_\mu,P_m,\Omega^\mu\\D_{\mu'},P_{m'},\Omega^{\mu'}\end{array}\right.
&\left\{
\begin{array}{l}
S_{mn},D_\mu,P_m,\Omega^\mu,\Sigma^{mn}\\
S_{m'n'},D_{\mu'},P_{m'},\Omega^{\mu'},
\Sigma^{m'n'}\end{array}\right.
\end{array}}
\eea  
For type II strings we double whole set of generators in the T-dual formalism.
A set of dimensional reduction constraints, coset constraints and the section condition are 
imposed to remove unphysical degrees of freedom in the end. 
 
\begin{itemize}
\item{Open superstring

Nondegenerate super-Poincar\'{e} covariant derivatives for a particle $\mathring{\nabla}_M$, the one 
 for a string $\mathring{\dd}_M$ and coordinates $Z^M$ are followings: 
\bea
{\renewcommand{\arraystretch}{1.6}
\begin{array}{cll}
&{\rm Indices}&_M=(_{mn},~_{\mu},~_{m},~^{\mu},~^{mn})\\
&{\rm Flat~supercovariant~derivative:}&
\mathring{\nabla}_M=(s,~d,~p,~\omega, ~\sigma)
\\
&{\rm Flat~super ~current:}&
\mathring{\dd}_M=(S,~D,~P,~\Omega,~\Sigma)\\
&{\rm Supercoordinate:}&Z^M=(u,~\theta,~x,~\varphi,~v)
\end{array}
\label{SDPWZ}}
\eea
 
  }
 
 \item{Type II superstring

 Now we double all currents and coordinates in order to
construct manifestly T-dual formulation of the type II theory.
Left and right currents  are 
denoted by unprimed indices and primed indices respectively:
\bea
{\renewcommand{\arraystretch}{1.6}
\begin{array}{ll}
{\rm Indices}&_M=(_{mn},_{\mu},_{m},^{\mu},^{mn}),~
_{M'}=(_{m'n'},_{\mu'},_{m'},^{\mu'},^{m'n'})
\\
{\rm Flat~supercovariant~derivative }:&\mathring{\dd}{}_{\underline{M}}=( \mathring{\dd}{}_M,~ \mathring{\dd}{}_{M'})
\\
~~~~~~~~~~~~~~~~~~~~~~~~~ {\rm flat~left}:&\mathring{\dd}{}_{M}=
(S_{mn},~D_\mu,~P_m,~\Omega^{\mu},~\Sigma^{mn})\\
~~~~~~~~~~~~~~~~~~~~~~~~~ {\rm flat~right}:&  \mathring{\dd}{}_{M'}=
(S_{m'n'},~D_{\mu'},~P_{m'},~\Omega^{\mu'},~\Sigma^{m'n'})\\
{\rm Supercoordinates}:&Z^{\underline{M}}=(Z^M,~Z^{M'})
\end{array}}\nn\\
\label{typeIIaffine}
\eea

Type IIA or type IIB is determined by the choice of two kinds of 
fermions.
In ten dimension the chiral representation is used as
\bea
&\left\{ \Gamma_m,\Gamma_n \right\}=2\eta_{mn}~~,~~
\Gamma_m=
\left(
\begin{array}{cc}
&(\gamma_m)^{\mu\nu}\\
(\gamma_m)_{\mu\nu}
\end{array}
\right)~~,~~
\Gamma_{mn}=
\left(
\begin{array}{cc}
(\gamma_{mn})^{\mu}{}_{\nu}&\\
&(\gamma_{mn})_{\mu}{}^{\nu}
\end{array}
\right)&\nn\\
&\Psi=\left(
{\renewcommand{\arraystretch}{1.4}
\begin{array}{c}\psi^\mu\\\chi_\mu\end{array}}
\right)~~,~~
\lbrack S_{mn},\Psi]=\frac{i}{2}
\left({\renewcommand{\arraystretch}{1.4}
\begin{array}{c}
(\gamma_{mn})^\mu{}_\nu\psi^\nu\\
(\gamma_{mn})_\mu{}^\nu\chi_\nu
\end{array}}
\right)~~~.&\nn
\eea
Type IIA/IIB fermions are assigned as
\bea
&{\rm type~IIA}&(Z^\mu,~Z^{\mu'})=(\theta^\mu,~\theta_\mu)\nn\\
&{\rm type~IIB}&(Z^\mu,~Z^{\mu'})=(\theta_1{}^\mu,~\theta_2{}^\mu)\nn~~~,
\eea
with respect to a common Lorentz symmetry generator
which is defined only after the dimensional reduction.
    }
\end{itemize}   
\par
\vskip 6mm
\subsection{Affine nondegenerate super-Poincar\'{e} algebra}

The affine nondegenerate super-Poincar\'{e} algebra
generated by \bref{typeIIaffine} is given by
\bea
\lbrack \mathring{\dd}{}_{\underline{M}}(1),\mathring{\dd}{}_{\underline{N}}(2)\}&=&
-if_{\underline{MN}}{}^{\underline{K}}\mathring{\dd}{}_{\underline{M}}\delta(2-1)
-i\eta_{\underline{MN}}\partial_\sigma\delta(2-1)
\eea
where structure constants and nondegenerate metrics  in left and right modes are
given as 
\bea
{\renewcommand{\arraystretch}{1.6}
\begin{array}{lcl}
\lbrack\mathring{\dd}{}_{M}(1),\mathring{\dd}{}_{N}(2) \}&=&
-if_{MN}{}^K
\mathring{\dd}{}_K\delta(2-1)-i\eta_{MN}\partial_{\sigma}\delta(2-1)
\nn\\
\lbrack\mathring{\dd}{}_{M'}(1),\mathring{\dd}{}_{N'}(2) \}&=&if_{MN}{}^K
\mathring{\dd}{}_{K'}\delta(2-1)+i\eta_{MN}\partial_{\sigma}\delta(2-1)~\nn\\
\lbrack\mathring{\dd}{}_{M}(1),\mathring{\dd}{}_{N'}(2)\}&=&0
\end{array}}
~~.\nn
\eea
Canonical dimensions of $(S,D,P,\Omega,\Sigma)$
are $0,\frac{1}{2},1,\frac{3}{2},2$ respectively.
The $\sigma$-derivative  $\partial_\sigma$ 
 in the Schwinger term carry canonical dimension 2
 (string tension is abbreviated).
The affine nondegenerate super-Poincar\'{e} algebra 
with $m=0,1,\cdots,9$ and $\mu=1,\cdots,16$ is given by
 \bea
{\renewcommand{\arraystretch}{1.6}
\begin{array}{llcl}
{\rm dim }~0:&\lbrack S_{mn}(1),S_{lk}(2)]&=&-i\eta_{[k|[m}S_{n]|l]}\delta(2-1)\\
{\rm dim }~\frac{1}{2}:&\lbrack S_{mn}(1),D_{\mu}(2)]&=&-\frac{i}{2}(D\gamma_{mn})_\mu\delta(2-1)\\
{\rm dim }~1:&\lbrack S_{mn}(1),P_{l}(2)]&=&-iP_{[m}\eta_{n]l}\delta(2-1)\\
&\left\{D_\mu(1),D_\nu(2)\right\}&=&2P_m \gamma^m{}_{\mu\nu}\delta(2-1)\\
{\rm dim }~\frac{3}{2}:&\lbrack S_{mn}(1),\Omega^\mu(2)]&=&
\frac{i}{2}(\gamma_{mn}\Omega)^\mu \delta(2-1)\\
&\lbrack D_\mu(1),P_n(2)]&=&2(\gamma_n\Omega)_\mu \delta(2-1)\\
{\rm dim }~2:&\lbrack S_{mn}(1),\Sigma^{lk}(2)]&=&-i\delta^{[k}_{[m}\Sigma_{n]}{}^{l]}\delta(2-1)
+i\delta_{[m}^l\delta_{n]}^k\partial_\sigma\delta(2-1)
\\
&\left\{D_\mu(1),\Omega^\nu(2)\right\}&=&
-\frac{i}{4}\Sigma^{mn}(\gamma_{mn})^\nu{}_\mu \delta(2-1)
+i\delta_\mu^\nu \partial_\sigma\delta(2-1)\\
&\lbrack P_{m}(1),P_{n}(2)]&=&i\Sigma_{mn}\delta(2-1)
+i\eta_{mn}\partial_\sigma\delta(2-1)
\end{array}}\label{SDPomesig}
\eea
Commutators with  dimension greater than 5/2 are zero.
The gamma matrices satisfy 
\bea
(\gamma^m)_{\mu\nu}=(\gamma^m)_{\nu\mu}~,~
(\gamma^{(m|})^{\mu\rho}(\gamma^{|n)})_{\rho\nu}=2\eta^{mn}\delta^\mu_\nu
~,~(\gamma_{m})_{(\mu\nu}(\gamma^m)_{\rho)\lambda}=0\nn~~~.
\eea
The right currents $(S',D',P',\Omega',\Sigma')$
satisfy the same algebra with opposite sign.

The nondegenerate metric $\eta_{\underline{MN}}$ is  denoted as:
\bea
&\begin{array}{lcl}
\eta_{\underline{MN}}&=&
\left(
\begin{array}{cc}
\eta_{MN}&0\\
0&\eta_{M'N'}
\end{array}\right)\\\\
&=&
\left(
\begin{array}{cc}
\eta_{MN}&0\\
0&-\eta_{MN}
\end{array}\right)\end{array}
,~
\eta_{{MN}}
=
\begin{array}{c}_S\\_D\\_P\\_\Omega\\_\Sigma\end{array}
\left(\begin{array}{ccccc}
& & & &\delta_{[m}^l\delta_{n]}^{k}\\
 & & &\delta_\mu^\nu& \\
 & &\eta_{mn}& & \\
 &-\delta_\nu^\mu& & & \\
\delta^{[m}_l\delta^{n]}_{k}& & & & 
\end{array}\right)~~&\label{etaMN}
\eea
\par
\vskip 6mm
\section{$\kappa$-symmetric Virasoro constraints}

The background of a bosonic
 string is determined by the Virasoro constraints.
The Green-Schwarz superstring has $\kappa$-symmetry 
which is necessary to eliminate a half of fermionic  degrees of freedom.
So the  background of the Green-Schwarz superstring 
is determined by  $\kappa$-symmetry covariant Virasoro constraints.
In this section a consistent set of the $\kappa$-symmetry constraints 
and the Virasoro constraints are obtained.

\vskip 6mm
\subsection{Virasoro constraints}
The Virasoro constraints are
 the Hamiltonian constraint ${\cal H}_\tau$ and the
$\sigma$ diffeomorphism constraint ${\cal H}_\sigma$,
which are written in bilinear of brane currents
to generate generalized gauge symmetries 
\cite{Hatsuda:2012vm,Hatsuda:2012uk}, as
\bea
{\cal H}_\tau&=&\displaystyle\frac{1}{2}\mathring{\dd}{}_{\underline{M}}
\hat{\delta}^{\underline{MN}} \mathring{\dd}{}_{\underline{N}}\label{Virasoro}\\
{\cal H}_\sigma&=&\displaystyle\frac{1}{2}\mathring{\dd}{}_{\underline{M}}
\eta^{\underline{MN}} \mathring{\dd}{}_{\underline{N}}\nn
\eea
$\hat{\delta}_{\underline{MN}}$ and $\eta^{\underline{MN}}$
are
\bea
\hat{\delta}^{\underline{MN}}&=&
\left(
\begin{array}{cc}
\eta^{MN}&0\\
0&\eta^{M'N'}
\end{array}\right)
=
\left(
\begin{array}{cc}
\eta^{MN}&0\\
0&\eta^{MN}
\end{array}\right)\nn\\\nn\\
\eta^{\underline{MN}}&=&
\left(
\begin{array}{cc}
\eta^{MN}&0\\
0&\eta^{M'N'}
\end{array}\right)
=
\left(
\begin{array}{cc}
\eta^{MN}&0\\
0& -\eta^{MN}
\end{array}\right)
\nn\\\nn\\
\eta^{MN}
&=&
\begin{array}{c}_S\\_D\\_P\\_\Omega\\_\Sigma\end{array}
\left(\begin{array}{ccccc}
& & & &\delta^{[m}_l\delta^{n]}_{k}\\
 & & &-\delta^\mu_\nu& \\
 & &\eta^{mn}& & \\
 &\delta^\nu_\mu& & & \\
\delta_{[m}^l\delta_{n]}^{k}& & & & 
\end{array}\right)\nn
\eea
satisfying  
\bea
{\eta}^{\underline{MN}}\eta_{\underline{NK}}{\eta}^{\underline{KL}}=
\hat{\delta}^{\underline{MN}}\eta_{\underline{NK}}\hat{\delta}^{\underline{KL}}=
\eta^{\underline{ML}}~~~.\label{etadeltaeta}
\eea

Virasoro algebra is given by 
\bea
{\renewcommand{\arraystretch}{1.6}
\begin{array}{ccl}
\lbrack
{\cal H}_\sigma(1),~{\cal H}_\sigma(2)]&=&
i\left({\cal H}_\sigma(1)+{\cal H}_\sigma(2)\right)
\partial_\sigma\delta(2-1)\\
\lbrack{\cal H}_\tau(1),~{\cal H}_\tau(2)]&=&
i\left({\cal H}_\sigma(1)+{\cal H}_\sigma(2)\right)
\partial_\sigma\delta(2-1)\\
\lbrack{\cal H}_\sigma(1),~{\cal H}_\tau(2)]&=&
i\left({\cal H}_\tau(1)+{\cal H}_\tau(2)\right)
\partial_\sigma\delta(2-1)
\end{array}}\label{Vira}
\eea
where \bref{etadeltaeta} is used.

The diffeomorphism constraint 
 ${\cal H}_\sigma$ generates  $\sigma$ derivative 
and $\mathring{\dd}_{\underline{M}}$ generates covariant derivatives as
\bea
&\partial_\sigma \Phi(Z^{\underline N})=
[\int i{\cal H}_\sigma,\Phi(Z^{\underline N})]~~,~~
[i\mathring{\dd}_{\underline M}(1),\Phi(Z^{\underline N}(2))]
=D_{\underline M}\Phi(Z^{\underline N})\delta(2-1)&~~~\nn\\
&~~\Rightarrow~~
\partial_\sigma \Phi(Z^{\underline N})=
\mathring{\dd}_{\underline M} 
\eta^{\underline MK}
\left(D_{\underline K}\Phi(Z^{\underline N})\right)~~.&\label{sigmader}
\eea

 Virasoro constraints
${\cal H}_\tau$ and ${\cal H}_\sigma$ 
 are separated into left and right  Virasoro
constraints ${\cal A}$ and ${\cal A}'$ respectively 
\bea
{\cal A}&=&\frac{1}{2}\left({\cal H}_\tau+{\cal H}_\sigma\right)
=\frac{1}{2}\mathring{\dd}_M\eta^{MN}\mathring{\dd}_N
\label{ABCDA}\\
{\cal A}'&=&
\frac{1}{2}\left({\cal H}_\tau-{\cal H}_\sigma\right)
=-\frac{1}{2}\mathring{\dd}_{M'}\eta^{MN}\mathring{\dd}_{N'}
\nn~~~.
\eea
\par
\vskip 6mm
\subsection{Fermionic constraints and $\kappa$-symmetry}

The covariant expression of the Green-Schwarz superstring has
 fermionic constraints, $D_{\underline{\mu}}=0$.
The $\kappa$-symmetry is generated by first class constraints
${\cal B}^\mu=D_{{\nu}}P_m(\gamma^m)^{\nu\mu}$
and ${\cal B}^{\mu'}=D_{{\nu}'}P_{m'}(\gamma^{m})^{\nu\mu}$ which is 
a half of $D_{\underline{\mu}}$.
Another half  is second class constraint.
Instead of imposing second class constraints,
first class constraints were constructed by bilinear of the second class constraints
 \cite{Siegel:1985xj} as  ${\cal C_{\mu\nu},D}_{m}=0$.

Two approaches of fermionic constraints are;
\begin{itemize}
  \item {Second class approach:
  \bea
{\rm  First~ class~ constraints}&:&
{\cal A}={\cal B}^\mu=0\nn\\
{\rm  Second~ class~ constraints}&:&    D_\mu=0\nn
  \eea}
  \item{First class approach:
  \bea
{\rm  First~ class~ constraints}&:&
{\cal A}={\cal B}^\mu={\cal C}_{\mu\nu}={\cal D}_m=0\nn
  \eea}
\end{itemize}
In this section we extend these conditions 
to the nondegenerate super-Poincar\'{e} case.

We extend ${\cal ABCD}$ constraints to the nondegenerate super-Poincar\'{e} space
as
 \bea
{\cal A}&=&\frac{1}{2}P_mP_n\eta^{mn}
+\Omega^\mu D_\mu+\frac{1}{2}\Sigma^{mn}S_{mn}\nn\\
{\cal B}^\mu&=&(D\gamma^m)^\mu P_m-iS_{mn}(\gamma^{mn}\Omega)^\mu
=(D\slP)^\mu-i(\slS\Omega)^\mu
\nn\\
{\cal C}_{\mu\nu}&=&D_\mu D_\nu
+\frac{1}{2i}S_{mn}P_{l}(\gamma^{mnl})_{\mu\nu}
=\frac{1}{2}D_{[\mu} D_{\nu ]}
+\frac{1}{4i}\left(\slS \slP\right)_{[\mu\nu]}
\nn\\
{\cal D}_{m}&=&(D\gamma_{m}\partial_\sigma D)+\frac{4}{i}
\Sigma_{mn}S^{nl}P_l
\label{ABCD}~~~.
\eea
Although ${\cal D}_m$ constraint includes trilinear term, 
 $\Sigma^{mn}$ commutes with other generators except $S$.

The ${\cal ABCD}$ algebra is computed analogously to the Virasoro constraints
by using the metric relation and antisymmetricity of the structure constant.
The ${\cal ABCD}$ constraints in the nondegenerate super-Poincar\'{e} space 
 given in \bref{ABCD} satisfy the following algebra;
\bea
{\renewcommand{\arraystretch}{1.6}
\begin{array}{lcl}
\lbrack{\cal A}(1),{\cal A}(2)]&=&i\left({\cal A}(1)+{\cal A}(2)\right)
\partial_\sigma\delta(2-1)\\
\lbrack{\cal A}(1),{\cal B}^\mu(2)]&=&i\left({\cal B}^\mu(1)+{\cal B}^\mu(2)\right)\partial_\sigma\delta(2-1)\\
\lbrack{\cal A}(1),{\cal C}_{\mu\nu}(2)]&=&i\left({\cal C}_{\mu\nu}(1)+{\cal C}_{\mu\nu}(2)\right)\partial_\sigma\delta(2-1)\\
\lbrack{\cal A}(1),{\cal D}_m(2)]&=&i\left({\cal D}_m(1)+2{\cal D}_m(2)\right)\partial_\sigma\delta(2-1)\\
\lbrack{\cal B}^\mu(1),{\cal B}^\nu(2)]&=&
i\frac{1}{2}
\left({\cal C}_{\rho\lambda}(1)+{\cal C}_{\rho\lambda}(2)\right)
(\gamma_m)^{\mu\rho}(\gamma^m)^{\nu\lambda}
\partial_\sigma\delta(2-1)\\
&&+\left[
(\gamma^m)^{\mu\nu}\left(4P_m{\cal A}+2({\cal B}\gamma_m\Omega)
+\frac{i}{2}{\cal D}_m
\right)\right.\\&&
\left.-4{\cal B}^{(\mu}\Omega^{\nu)}
+i{\cal C}_{\rho\lambda}(\gamma^m)^{\mu\rho}(\gamma^n)^{\nu\lambda}
\Sigma_{mn}
\right]\delta(2-1)\\
\lbrack{\cal B}^\mu(1),{\cal C}_{\nu\rho}(2)]&=&
\left[ 4{\cal A}\delta^\mu_{[\nu}D_{\rho]}
+\frac{1}{8}{\cal B}^{\lambda}(\gamma_{mn})_\lambda{}^\mu(\gamma^{mn}\slP)_{[\nu\rho]}
\right.\\
&&+{\cal C}_{\sigma[\nu}c^{\mu\sigma}_{\rho]}
\left.+S_{mn}\tilde{c}^{mn;\mu}_{\nu;\rho}
-(\partial_\sigma S_{mn})\frac{1}{2}(D\gamma_l)^\mu(\gamma^{mnl})_{\nu\rho}
  \right]\delta(2-1)
\\
&&+S_{mn}\left(
\frac{1}{2}(D\gamma_l)^\mu(\gamma^{mnl})_{\nu\rho}
-(\gamma^{mn})^\mu{}_{[\nu}D_{\rho]}
\right)(1)\partial_\sigma\delta(2-1)
\\
\lbrack {\cal B}^\mu(1),{\cal D}_m(2)]&=&
\left[
-4{\cal A}(D\gamma_m)^\mu+{\cal C}_{\rho\nu}(\gamma^n)^{\nu\mu}(\gamma_n\gamma_m\Omega)^\rho
\right](1)\partial_\sigma\delta(2-1)\\
&&+\left[
8{\cal A}\partial_\sigma(D\gamma_m)^\mu
-2({\cal B}\gamma^{nl})^\mu\Sigma_{mn}P_l\right.\\
&&\left.
+4\partial_\sigma{\cal C}_{\nu\rho}\left((\gamma_m)^{\mu\nu}\Omega^\rho
+\frac{1}{2}(\gamma^n)^{\mu\nu}
(\gamma_m\gamma_n\Omega)^\rho\right)\right.\\
&&\left.
+2{\cal D}^n(\gamma_n\gamma_m\Omega)^\mu
+2{\cal D}^l(\gamma_{lm}\Omega)^\mu+c^\mu{}_m{}^{nl}S_{nl}
\right]\delta(2-1)\\
\lbrack{\cal C}_{\mu\nu}(1),{\cal C}_{\rho\lambda}(2)]&=&
-\frac{i}{16}(\slS \gamma^l)_{[\mu\nu]}
(\slS \gamma_l)_{[\rho\lambda]}(1)\partial_\sigma\delta(2-1)\\&&
+\lbrack{\cal C}_{\sigma\eta} c^{\sigma\eta}+S_{mn}\tilde{c}^{mn}_{\mu\nu;\rho\lambda}
]
\delta(2-1)\\
\lbrack {\cal C}_{\mu\nu}(1),{\cal D}_m(2)]&=&
-4{\cal C}_{\mu\rho}(\slP \gamma_m)_\nu{}^\rho(1)\partial_\sigma\delta(2-1)\\
&&+\left[-4({\cal B}\gamma_m)_{[\mu}\partial_\sigma D_{\nu]}
+8(\partial_\sigma{\cal C}_{\mu\nu})P_m
-8(\partial_\sigma{\cal C}_{\mu\rho})(\gamma \slP)_\nu{}^\rho\right.\\
&&\left.
-2{\cal C}_{\rho[\mu}(\gamma^{nl})^\rho{}_{\nu]}\Sigma_{mn}P_l
-\frac{1}{2}{\cal D}^{l}(\gamma_{lm}\slP)_{\mu\nu}+c_{\mu\nu m}^{nl}S_{nl}
\right]\delta(2-1)\\
\lbrack {\cal D}_{m}(1),{\cal D}_n(2)]&=&
-2{\cal C}_{\mu\nu}(\slP\gamma_{mn})^{\mu\nu}(1)\partial_\sigma{}^2(2-1)\\
&&+\left(
2{\cal C}_{\mu\nu}\partial_\sigma(\slP \gamma_{mn})^{\mu\nu}

-8{\cal D}_{(m}P_{n)}+8{\cal D}_lP^l\eta_{mn}
\right)\partial_\sigma\delta(2-1)\\&&
-4\Bigl[
\partial_\sigma({\cal B}\gamma_{mn}\partial_\sigma D)
+2(\partial_\sigma{\cal B})\gamma_{mn}(\partial_\sigma D)
-(\partial_\sigma{\cal C}_{\mu\nu})(\slP \gamma_{mn})^{\mu\nu}\\
&&\left.-2{\cal D}^k\Sigma^{m[k}P_{n]}-\partial_\sigma({\cal D}_{[m}P_{n]})
+c_{mn}^{lk}S_{lk}
\right]\delta(2-1)
\end{array}
}\label{ABCDalg}
\eea
where  $c^\mu{}_m{}^{nl}$, $c_{\mu\nu m}{}^{nl}$
and $c_{mn}{}^{lk}$ are  coefficient functions, for example
\bea
\lbrack{\cal B},{\cal C}]&\ni&\nn\\
~~~~~~~~~c^{\mu\sigma}_{\rho}&=&
4\delta_{\rho}^\mu\Omega^\sigma
-2(\gamma^m)^{\mu\sigma}(\gamma_m\Omega)_{\rho}
+\frac{1}{2}(\gamma_{mn})^\sigma{}_{\rho}(\gamma^{mn}\Omega)^\mu\nn\\
~~~~~~~~~\tilde{c}^{mn;\mu}_{\nu;\rho}&=&
-\frac{1}{2i}(\gamma^{mnl})_{\sigma[\nu}P_l(c^{\mu\sigma}_{\rho]}
-2\delta^\sigma_\rho\Omega^\mu)
-2\Sigma^{mn}\delta^\mu_{[\nu}D_{\rho]}\nn\\
&&-\frac{1}{4}(\gamma^{mn}\slSigma)^\mu{}_{[\nu}D_{\rho]}
+\frac{1}{2}(D\gamma^k)\Sigma_{kl}(\gamma^{mnl}_{\nu\rho})
+(\gamma^{mn})^\mu{}_{[\nu}\partial_\sigma D_{\rho]}\nn\\
\lbrack{\cal C},{\cal C}]&\ni&\nn\\
~~~~~~~~~c^{\sigma\eta}_{\mu\nu;\rho\lambda}&=&
2\delta^\eta_{[\lambda|}\delta^{\sigma}_{[\mu}\slP_{\nu]|\rho]}
+\frac{1}{8}\delta^{\sigma}_{[\mu}(\gamma_{mn})^\eta{}_{\nu]}
(\gamma^{mn}\slP)_{[\rho\lambda]}
-\frac{1}{8}\delta^{\sigma}_{[\rho}(\gamma_{mn})^\eta{}_{\lambda]}
(\gamma^{mn}\slP)_{[\mu\nu]}\nn\\
~~~~~~~~~\tilde{c}^{mn}_{\mu\nu;\rho\lambda}&=&
\frac{i}{4}c^{\sigma\eta}_{\mu\nu;\rho\lambda}(\gamma^{mn}\slP)_{[\sigma\eta]}
-iD_{[\mu}(\gamma_l\Omega)_{\nu]}(\gamma^{mnl})_{\rho\lambda}
+iD_{[\rho}(\gamma_l\Omega)_{\lambda]}(\gamma^{mnl})_{\mu\nu}\nn\\&&
-\frac{i}{4}(\gamma^{ml}\slP)_{[\mu\nu]}(\gamma^{n}{}_{l}\slP)_{[\rho\lambda]}
-\frac{i}{8}(\gamma^{mnl})_{\mu\nu}
\left(
(\gamma^k\slS)_{[\rho\lambda]}\Sigma_{lk}
-(\gamma_l\partial_\sigma\slS)_{[\rho\lambda]}\right)\nn
\eea
Since coefficients of $S_{mn}$ do not vanish even on the constrained surface,
consistent space is coset space G/H where G is nondegenerate super-Poincar\'{e} 
group and H is Lorentz group. Subgroup H is gauged and the coset space is
defined by
\bea
S_{mn}=0~~~.\label{cosetS}
\eea
${\cal ABCD}$ constraints in \bref{ABCD} are reducible
up to $S=0$ constraint ($\approx$ ) 
\bea
D_\mu{\cal B}^\nu \approx {\cal C}_{\mu\rho}\slP^{\rho\nu}~~,~~
({\cal B}\slP)_\mu\approx 2{\cal A}D_\mu-2{\cal C}_{\mu\rho}\Omega^\rho~~{\rm etc.}\nn
\eea
The type II theory  has two sets of constraints 
\bref{ABCD},  $({\cal ABCD}{\cal A'B'C'D'})=0$.
${\cal A'B'C'D'}$ are 
constructed by bilinears of right currents $\mathring{\dd}{}_{M'}$.
They satisfy the same algebra \bref{ABCDalg}
with opposite sign, and ${\cal ABCD}$ commute with ${\cal A'B'C'D'}$.

Next the second class approach is also examined.
The consistency condition is the closure of algebra between
first and second class constraints $D_\mu=0$;
\bea
{\renewcommand{\arraystretch}{1.6}
\begin{array}{lcl}
\lbrack{\cal A}(1),D_\mu(2)]&=&iD_\mu(1)\partial_\sigma\delta(2-1)~\approx ~0
\nn\\
\left\{{\cal B}^\mu(1),D_{\nu}(2)\right\}&=&
\left[ 4{\cal A}\delta_\nu^\mu
-\left(
4\Omega^\rho \delta_\nu^\mu-(\gamma^m)^{\mu\rho}(\gamma_m\Omega)_\nu
+\frac{1}{2}(\gamma_{mn}\Omega)^\mu(\gamma_{mn})^\rho{}_\nu
\right)D_\rho\right.\nn\\&&\left.
+\left(2\Sigma^{nm}\delta_\nu^\mu+\frac{1}{4}(\gamma^{nm}\slSigma)^\mu{}_\nu
\right)S_{mn}
\right]\delta(2-1)\nn\\
&&-\slS^\mu{}_\nu
(1)\partial_\sigma\delta(2-1)~\approx ~0
\end{array}}~~\nn~~.
\eea
They vanish 
up to second class and first class constraints
 $D_\mu={\cal A}={\cal B}^\mu=S_{mn}=0$ ($\approx$ ).

Therefore the constraint sets for the type II Green-Schwarz superstring 
in  nondegenerate super-Poincar\'{e} space are summarized as;
\begin{itemize}
  \item {Second class approach: 
  \bea
{\rm First ~class~constraints}&:&~~{\cal A,B}^\mu={\cal A',B}^{\mu'}=S_{mn}=S_{m'n'}=0~~\nn\\
{\rm Second ~class~constraints}&:&~~D_{\mu}=D_{\mu'}=0~~\nn
\eea}
  \item{First class approach:
\bea
{\rm First ~class~constraints}:~~{\cal A,B^\mu,C_{\mu\nu},D}_m
={\cal A',B^{\mu'},C_{\mu'\nu'},D}_{m'}=S_{mn}=S_{m'n'}=0~~~\nn
\eea}
\end{itemize}

It is important that  
all the $\kappa$-symmetric
Virasoro constraints 
 are written by bilinears of $\mathring{\dd}$
  with arbitrary coefficients $a$'s
\bea
&
{\cal B}^\mu a_\mu=\frac{1}{2}\mathring{\dd}_M\rho^{MN}\mathring{\dd}_N~,~
{\cal C}_{\mu\nu}\check{a}^{\mu\nu}=\frac{1}{2}\mathring{\dd}_M
\check{\rho}^{MN}\mathring{\dd}_N~,~
{\cal D}_{m}\tilde{a}^{m}=\frac{1}{2}\mathring{\dd}_M
\tilde{\rho}^{MN}\mathring{\dd}_N\label{rhoBCD}
&\eea
\bea
\rho^{MN}&=&
\begin{array}{c}
_S\\_D\\_P\\_\Omega\\_\Sigma
\end{array}
\left(\begin{array}{ccccc}
&&&-i(\gamma^{mn})^{\mu}{}_{\nu}a_\mu&\\
&&(\gamma^n)^{\mu\nu}a_\nu&&\\
&(\gamma^m)^{\mu\nu}a_\mu&&&\\
-i(\gamma^{nl})^{\nu}{}_{\mu}a_\nu&&&&\\
&&&&
\end{array}
\right)~\nn\\
\check{\rho}^{MN}&=&
\begin{array}{c}
_S\\_D\\_P\\_\Omega\\_\Sigma
\end{array}
\left(\begin{array}{ccccc}
&&\frac{1}{2i}(\gamma^{mnl})_{\mu\nu}\check{a}^{\mu\nu}&&\\
&\check{a}^{[\mu\nu]}&&&\\
\frac{1}{2i}(\gamma^{nlm})_{\mu\nu}\check{a}^{\mu\nu}&&&&\\
&&&&\\
&&&&
\end{array}
\right)\nn\\
\tilde{\rho}^{MN}&=&
\begin{array}{c}
_S\\_D\\_P\\_\Omega\\_\Sigma
\end{array}
\left(\begin{array}{ccccc}
&&\frac{4}{i}\tilde{a}^k\Sigma_{km}\delta_n^l&&\\
&\tilde{a}^m(\gamma_n)^{(\mu\nu)}\partial_\sigma&&&\\
\frac{4}{i}\tilde{a}^k\Sigma_{kn}\delta_l^m&&&&\\
&&&&\\
&&&&
\end{array}
\right)
\nn
\eea
$\rho$'s are nilpotent matrices; 
$\rho^5=\check{\rho}^3=\tilde{\rho}^3=0$
by lowering index with $\eta_{MN}$.
They satisfy $\rho^2=\check{\rho}$, where this relation
gives $\{{\cal B},{\cal B}\}\approx {\cal C}
\partial_\sigma\delta$.
Sum of ${\cal ABCD}$ ${\cal A'B'C'D'}$  becomes a
manifestly T-dual bilinear as
\bea
\Xi(a)=\frac{1}{2}\mathring{\dd}_{{M}}\Xi^{{MN}}(a)\mathring{\dd}_{{N}}~~,~~
\Xi^{{MN}}(a)=
\left[\eta+\rho(a)+\check{\rho}(\check{a})+\tilde{\rho}(\tilde{a})
\right]^{{MN}}~~~.\label{Xi}
\eea 
 $\Xi$ is 
covariant under the T-duality rotation;
$\Xi(a)\to\Xi(a')=M\Xi(a)M^T$ under  
$\mathring{\dd} \to M\mathring{\dd}$ with
$M \eta M^T=\eta$,   since $\Xi^{{MN}}$ is upper triangular matrix. 
A set of $\kappa$-symmetric 
Virasoro constraints is consistent to T-duality symmetry.

\par
\vskip 6mm

\section{Manifestly T-dual formulation  
of type II superstring curved background}

We present a manifestly T-dual formulation of 
the low energy effective gravity theory for the type II superstring.
The procedure was given in \cite{Polacek:2013nla},
and we extend this to a type II superspace.
The space is spanned by the nondegenerate super-Poincar\'{e} algebra,
and curved background fields are expanded by this basis.
The vielbein includes R-R field strength
as well as all background fields.
Torsions include supercurvature tensors.
It turns out that all superconnections are solved in terms of 
a prepotential which is included in the vielbein.

\subsection{Vielbein, torsion and Bianchi identity}

Covariant derivatives  in curved background are 
 given with vielbein 
$E_{\underline{A}}{}^{\underline{M}}$ as
\bea
{\dd}_{\underline{A}}=
E_{\underline{A}}{}^{\underline{M}}(Z^{\underline N})\mathring{\dd}_{\underline{M}}
\label{DDED}
\eea
which satisfy the following algebra
\bea
 [{\dd}_{\underline{A}}(1),{\dd}_{\underline{B}}(2)\}&=&
-iT_{\underline{AB}}{}^{\underline{C}}{\dd}_{\underline{C}}\delta(2-1)
-i\eta_{\underline{AB}}\partial_\sigma\delta(2-1)~~~.\label{dddd}
\eea
Vectors in this space, $\hat{\Lambda}_i=\Lambda_i{}^{\underline{M}}\mathring{\dd}_{\underline{M}}$ with $i=1,2$, satisfy the following supersymmetric Lie bracket
in a manifestly T-dual formulation 
\bea
&&\left[\hat{\Lambda}_1,\hat{\Lambda}_2\right]_{\rm T}
=-i\hat{\Lambda}_{12}\nn\\
&&\Lambda_{12}{}^{\underline{M}}=\Lambda_{[1}{}^{\underline{N}}
(D_{\underline{N}}\Lambda_{2]}{}^{\underline{M}})
+\Lambda_{1}{}^{\underline{N}}(D^{\underline{M}}\Lambda_2{}_{\underline{N}})
+\Lambda_1{}^{\underline{N}}\Lambda_2{}^{\underline{L}}f_{\underline{NL}}
{}^{\underline{M}}~~~.\label{superC}
\eea
There is an ambiguity from Schwinger term proportional to
 $\partial_\sigma\delta$ in \bref{dddd}.  
The vielbein field $\hat{E}_{\underline{A}}=E_{\underline{A}}{}^
{\underline{M}}\mathring{\dd}_{\underline{M}}$ 
has gauge symmetry generated by the above bracket as
\bea
\delta_\Lambda \hat{E}_{\underline{A}}&=&
\left[\hat{E}_{\underline{A}},\hat{\Lambda}\right]_{\rm T}\nn~~~\\
\delta_\Lambda {E}_{\underline{A}}{}^{\underline{M}}&=&
{E}_{\underline{A}}{}^{\underline{N}}
(D_{\underline{N}}\Lambda^{\underline{M}})
-\Lambda{}^{\underline{N}}
(D_{\underline{N}}{E}_{\underline{A}}{}^{\underline{M}})
+{E}_{\underline{A}}{}^{\underline{N}}
(D^{\underline{M}}\Lambda_{\underline{N}})
+{E}_{\underline{A}}{}^{\underline{N}}\Lambda^{\underline{L}}f_{\underline{NL}}
{}^{\underline{M}}~~~.
\eea
The general coordinate transformation parameter is 
$\lambda^m+\lambda^{m'}$, the gauge transformation of $B$ field
is $\lambda^m-\lambda^{m'}$, the local 
supersymmetry parameters are $\lambda^\mu,~\lambda^{\mu'}$
and the two local Lorentz parameters are
 $\lambda^{mn}$ and $\lambda^{m'n'}$.
Other parameters are unphysical
whose symmetries are
fixed by dimensional reduction constraints 
in terms of $\tilde{\Sigma}$ and $\tilde{\Omega}$.

Manifestly T-dual formulation allows to 
impose orthonormal condition on vector parts of vielbein 
 $e_{\underline{a}}{}^{\underline{m}}=\left(\begin{array}{cc}
e_a{}^m&e_a{}^nB_{nm}\\
0&e_m{}^a
\end{array}
\right)$ for
$\mathring{\dd}_{\underline{m}}=(P_m,~\partial_\sigma x^m)$
which 
satisfies $e\eta e^T=\eta$
with $\eta_{MN}=\eta_{AB}=\left(
\begin{array}{cc}
&1\\1&
\end{array}
\right)$.
The vielbein field is an element of O(d,d) symmetry
which is continuous symmetry of zero-mode part of whole 
string T-duality.
For supersymmetric case it is enlarged to 
\bea
E_{\underline{A}}{}^{\underline{M}}\eta_{\underline{MN}}
E_{\underline{B}}{}^{\underline{N}}=\eta_{\underline{AB}}~~,\label{ortho}~~
\eta_{\underline{MN}}=\eta_{\underline{AB}}~~,
\eea
which is OSp(d$^2$,d$^2$$\mid$$2^{{\rm d}/2},2^{{\rm d}/2}$) for d dimensional space.
Both  flat currents  and curved currents satisfy algebras with 
the same Schwinger term. 
Torsion with lower indices is totally graded antisymmetric,
and is expressed from
 \bref{sigmader} and \bref{ortho} as
\bea
T_{\underline{ABC}}\equiv T_{\underline{AB}}{}^{\underline{D}}\eta_{\underline{DC}}
=\displaystyle\frac{1}{3!}T_{\underline{[ABC)}}
=\frac{1}{2}(D_{[{\underline{A}}}
E_{{\underline{B}}}{}^{\underline{M}})
E_{\underline{C})\underline{M}}
+E_{\underline{A}}{}^{\underline{M}}E_{\underline{B}}{}^{\underline N}E_{\underline C}{}^{\underline L}f_{\underline{MNL}}
~~~
\eea
with $D_{\underline{A}}=E_{\underline{A}}{}^{\underline{M}}D_{\underline{M}}$.
Graded antisymmetricity requires suitable sign factor which are omitted
in this expression,
with ${\cal O}_{[AB)}={\cal O}_{AB}-(-)^{AB}{\cal O}_{BA}$.
Bianchi identity gives the following totally graded antisymmetric tensor to vanish
\bea
{\cal I}_{\underline{ABCD}}&=&
\frac{1}{4!} {\cal I}_{[\underline{ABCD})}=
(D_{[{\underline A}}T_{{\underline{BCD}})})
+\frac{3}{4}T_{[{\underline AB}}{}^{\underline E}
T_{{\underline CD}){\underline E}}= 0~~~.
\eea

We divide the coset space G/H as follows;
\bea
{\renewcommand{\arraystretch}{1.3}
\left\{\begin{array}{l}
{\rm G}=(\underline{S},~\underline{D},~\underline{P},~
\underline{\Omega},~\underline{\Sigma})\\
{\rm H}=(\underline{S})\\
\tilde{\rm H}=(\underline{\Sigma})\\
{\rm K}=(\underline{D},~\underline{P},~
\underline{\Omega})\\
\end{array}\right.}
\eea
The torsion constraints with Lorentz lower indices are imposed 
in such a way that local Lorentz symmetries
satisfy the same algebra to the flat algebra 
\bea
T_{{\underline{SAB}}}=f_{{\underline{SAB}}}~~~.\label{TSTS}
\eea
The algebras among covariant derivatives in K, H, $\tilde{\rm H}$
are given as follows
with 
$\nabla_{I} \in$K$=(\underline{D},\underline{P},\underline{\Omega})$
\bea
{\renewcommand{\arraystretch}{1.8}
\left\{
\begin{array}{l}
\lbrack \nabla_I,\nabla_J\}=-iR_{IJ}{}^{\underline{S}}\nabla_{\underline{S}}
-iT_{IJ}{}^K \nabla_K -if_{IJ\underline{S}} \nabla_{\underline{\Sigma}}\\
\lbrack \nabla_{\underline{S}},\nabla_{\underline{S}} ]=
-if_{\underline{SS\Sigma}} \nabla_{\underline{S}} ~,~
\lbrack \nabla_{\underline{S}},\nabla_{\underline{\Sigma}}]=
-if_{\underline{S\Sigma S}} \nabla_{\underline{\Sigma}}~,~
\lbrack \nabla_{\underline{S}},\nabla_I]=
-if_{\underline{S}I}{}^{J} \nabla_{J}~~\\
\lbrack \nabla_{\underline{\Sigma}},\nabla_I]=
-iT_{\underline{\Sigma}I\underline{\Sigma}} \nabla_{\underline{S}}
-iR_{I}{}^{J\underline{S}} \nabla_{J}~,~
\lbrack \nabla_{\underline{\Sigma}},\nabla_{\underline{\Sigma}}]=
-iT_{\underline{\Sigma}\underline{\Sigma}{\underline{\Sigma}}} \nabla_{\underline{S}}
-iT _{{\underline{\Sigma}}{\underline{\Sigma}}}{}^{I} \nabla_{I}
\end{array}\right.}~~.\label{covKHH}
\eea
Nontrivial torsions are obtained only from $\lbrack \nabla_I,\nabla_J\}$.
Torsions with Lorentz upper index are curvature tensors. 
Supercurvatures $R_{IJ}{}^{\underline{S}}$ 
are  written in terms of superconnections 
$\omega_{I}{}^{\underline{S}}=E_{I}{}^{\underline{S}}$ as
\bea
R_{IJ}{}^{\underline{S}}=T_{IJ\underline{\Sigma}}&=&
-(D_{I}E_J{}^{\underline{M}})\omega_{\underline{M}}{}^{\underline{S}}
-(D_{J}\omega_{\underline{M}}{}^{\underline{S}})
E_{I}{}^{\underline{M}}
+(D_{\Sigma}E_{\underline{I}}{}^{\underline{M}})
E_{J}{}_{\underline{M}}
\nn\\
&&-(\omega_I{}^{\underline{S}}E_J{}^{\underline{M}}
\omega^{\underline{N}}{}^{\underline{S}}
+E_I{}^{\underline{M}}E_J{}^{\underline{N}}r^{\underline{S}}{}_{\underline\Sigma}
+E_I{}^{\underline{N}}\omega_J{}^{\underline{S}}
\omega^{\underline{M}}{}^{\underline{S}}
)f_{\underline{SMN}}\nn\\
&&-(E_I{}^{\underline{\alpha}}E_J{}^{\underline{\beta}}
\omega^{\underline{P}}{}^{\underline{S}}
+E_I{}^{\underline{P}}E_J{}^{\underline{\alpha}}
\omega^{\underline{\beta}}
{}^{\underline{S}}
+E_I{}^{\underline{\beta}}E_J{}^{\underline{P}}
\omega^{\underline{\alpha}}{}^{\underline{S}})
f_{\underline{\alpha\beta P}}
~~~.
\eea
Some curvature tensors are related by Bianchi identities,
and 
$R_{mn}{}^{\underline{S}}=\frac{1}{2}(\gamma_{mn})^{\mu}{}_{\nu}
R_{\mu}{}^{\nu \underline{S}}$.
Torsions with two and three $\underline{\Sigma}$ lower indices, 
$T_{\underline{\Sigma\Sigma}I}$
and $T_{\underline{\Sigma\Sigma\Sigma}}$,
are unphysical.

\par\vskip 6mm
 
\subsection{Torsion constraints from $\kappa$-symmetric Virasoro}

Let us obtain conditions on torsions from
the Green-Schwarz superstring in curved space. 
Virasoro constraints in curved space are 
\bref{Virasoro} by replacing $\mathring{\dd}_{\underline{M}}$ with
${\dd}_{\underline{A}}$ as
\bea
{\cal H}_\tau&=&\frac{1}{2}\dd_{\underline{A}}\hat{\delta}^{\underline{AB}}
\dd_{\underline{B}}
\label{curvedVira}\\
{\cal H}_\sigma&=&\frac{1}{2}\dd_{\underline{A}}{\eta}^{\underline{AB}}
\dd_{\underline{B}}
=\frac{1}{2}\mathring{\dd}_{\underline{M}}{\eta}^{\underline{MN}}
\mathring{\dd}_{\underline{N}}~~~.
\nn
\eea 
They satisfy the same  Virasoro algebra as in \bref{Vira} 
from totally antisymmetricity of torsion and the relations of metric
$\eta$ and $\hat{\delta}$ without introducing  new torsion constraints. 
The ``section condition" is the
 $\sigma$-diffeomorphism constraint ${\cal H}_\sigma=0$
  which guarantees geometry generated by the bracket \bref{superC}
from \bref{sigmader}.

Natural extension of ${\cal ABCD}$ and
${\cal A'B'C'D'}$ constraints in \bref{ABCD} and \bref{rhoBCD}
to the ones in curved space  are
obtained by replacing $\mathring{\dd}_{\underline{M}}$ 
with $\dd_{\underline{A}}$ with the same matrices $\rho$'s in \bref{rhoBCD}.
It is obtained by replacing  $\mathring{\dd}=E^{-1}{\dd}$
in ${\cal ABCD}$ constraints from the relation   
in \bref{Xi}.
Sets of constrains, which determine curved background fields of 
the type II Green-Schwarz superstring, are the followings; 
\begin{itemize}
  \item {Second class approach:
\bea
\begin{array}{lcl}
{\rm First~class~constraints}&:&
{\cal A,B}^\alpha={\cal A',B}^{\alpha'}={\dd}_{\underline{S}}= 0~~~\\
{\rm Second~class~constraints}&:&{\dd}_{D_{\underline{\alpha}}}
= 0~~
\end{array}\nn
\eea
 }
 \item{First class approach:
 \bea
\begin{array}{lcl}
{\rm First~class~constraints}&:&
{\cal A,B^\alpha,C_{\alpha\beta},D}_a={\cal A',B^{\alpha'},C_{\alpha'\beta'},D}_{a'}
={\dd}_{\underline{S}}= 0~~~.
\end{array}\nn
\eea   }
\end{itemize}
Both notations of the fermionic covariant derivative, 
${\dd}_{D_{\underline{\alpha}}}={\dd}_{\underline{\alpha}}$,
are used.

We analyze both two approaches of fermionic constraints 
analogously  to \cite{Shapiro:1986xp}.
Let us examine the second class approach.
Commutator of the second class constraints and the diffeomorphism constraint
is trivial,  $[{\dd}_{\underline{\alpha}},\int {\cal H}_\sigma]=i\partial_\sigma \dd_{\underline{\alpha}}\approx 0$. 
Commutator of ${\dd}_{{\underline{\alpha}}}$ 
 with ${\cal H}_\tau$ gives following conditions
\bea
\lbrack{\dd}_{\underline{\alpha}},{\cal H}_\tau
]&=&-i
T_{\underline{\alpha BC}}( \eta^{\underline{BD}} {\dd}_{\underline{D}})
(\hat{\delta}^{\underline{CE}} {\dd}_{\underline{E}})
\delta(2-1)
+i(-)^{\underline{\alpha}}{\dd}_{\underline{\alpha}}(1)
\partial_\sigma\delta(2-1)\nn\\
&\approx&0\nn\\
&\Leftrightarrow
&~
T_{\underline{\alpha} BC'}
=0~~~{\rm for}~_{B,C=(S,D_{\alpha},P)}\label{Torsion1}
\eea
These constraints $T_{\underline{\alpha}B{C}'}=0~$ reflect holomorphicity.

Further constraints are obtained from  anticommutator with the 
$\kappa$-symmetry 
generator
\bea
&&\{{\dd}_{\underline{{\alpha}} },{\cal B}^{\underline{\beta}}\}
\approx 0~\nn\\
&&~~~~~~\Leftrightarrow
\left\{{\dd}_\alpha,{\dd}_\beta \right\}\approx f_{\alpha\beta c}{\dd}_{c}~,~
\left\{{\dd}_{\alpha'},{\dd}_{\beta'} \right\}\approx -f_{\alpha\beta c}{\dd}_{c'}~,~
\left\{{\dd}_\alpha,{\dd}_{\beta'} \right\}\approx 0\nn\\
&&~~~~~\Rightarrow 
T_{\alpha \beta c}=f_{\alpha\beta c},~ 
T_{{\alpha\beta}X}=0 ~{\rm for}~_{X=(S,S',D_{{\alpha}},D_{{\alpha}}',P')}
~~,~~{\rm same ~for~{}'{}~indices}~
\eea
where $\approx$ allows terms including 
$\dd_{{\underline{\alpha}}}$ and $\dd_{S}$.
Bianchi identity  and totally graded antisymmetricity restrict torsion as
$T_{\alpha\beta}{}^{\underline{\gamma}}=0
=T_{\alpha a}{}^{\underline{\gamma}}$.
Torsion constraints are 
\bea
{\renewcommand{\arraystretch}{1.6}
\left\{
\begin{array}{l}
T_{\underline{\alpha} BC'}
=0~~{\rm for}~_{B,C=(S,D_{\alpha},P)}\\
T_{\alpha \beta c}=f_{\alpha\beta c }
~,~T_{{\alpha\beta}X}=0 ~{\rm for}~_{X=(S,S',D_{{\alpha}},D_{{\alpha}'},P',\Omega,\Omega')}~~\\
T_{{\alpha b}Y}=0 ~
{\rm for}~_{Y=(S,S',D_{{\alpha}'},P,P',\Omega,\Omega')}\\
~~{\rm same ~for~'~indices}~.~
\end{array}\right.}~\label{2Torsion}
\eea

In the first class approach 
 commutators among ${\cal ABCD}$ constraints 
lead to the same conditions to the above
up to $\dd_S$ terms.
Commutators 
     $\lbrack{\cal B}^{\underline{\alpha}},{\cal H}_\tau]$
and 
 $\{{\cal B}^{\underline{\alpha}},{\cal B}^{\underline{\beta}}\}$ 
 lead to the same torsion condition \bref{2Torsion}.
 \par
 \vskip 6mm
 
\subsection{Prepotential}

The consistency condition of the nondegenerate super-Poincar\'{e} space
leads to Lorentz constraint, \bref{cosetS},
so it is possible to choose a local Lorentz frame as
\bea
\dd_{S_{\underline{ab}}}=S_{\underline{ab}}~~~.
\eea 
This space is a coset space G/H where H is Lorentz,
then H coordinates can be removed without modifying 
algebra of covariant derivatives supplying H constraints.
This is because the symmetry generator satisfies
the symmetry algebra regardless of the presence or absence of H-coordinates,
and the algebra of symmetry generators and covariant derivatives 
have same structure constant with opposite sign.
Covariant derivatives include $\Sigma$ terms which can be removed
with supplying $S$ constraints,
so the coefficient of $\Sigma$ in vielbein can be chosen to zero.   
Vielbein fields can be parameterized as 
\bea
&&\nn\\
&&E_{{\underline{A}}}{}^{\underline{M}}=
{\renewcommand{\arraystretch}{1.6}
\begin{array}{c}
_{S_{\underline{ab}}}\\_{D_{\underline{\alpha}}}\\_{P_{\underline{a}}}\\_{\Omega^{\underline{\alpha}}}\\_{\Sigma^{\underline{ab}}}
\end{array}
\left(
\begin{array}{ccccc}
{\bf 1}_{\underline{ab}}{}^{\underline{mn}}&0&0&0&0\\
\omega_{\underline{\alpha}}{}^{\underline{mn}}&e_{\underline{\alpha}}{}^{\underline{\mu}}&
{B}_{\underline{\alpha}}{}^m&B_{\underline{\alpha\mu}}&0\\
\omega_{\underline{a}}{}^{\underline{mn}}&\psi_{\underline{a}}{}^{\underline{\mu}}
&e_{\underline{a}}{}^{\underline{m}}&
-\check{B}_{\underline{a\mu}}&0\\
\omega^{\underline{\alpha}}{}^{\underline{mn}}&
F^{\underline{\alpha\mu}}&-\check{\psi}^{\underline{\alpha}}{}^{\underline{m}}&e^{\underline{\alpha}}{}_{\underline{\mu}}&0\\
r^{\underline{abmn}}&-\check{\omega}^{\underline{ab}}{}^{\underline{\mu}}&
-\check{\omega}^{\underline{ab}}{}^{\underline{m}}&-\check{\omega}^{\underline{ab}}{}
_{\underline{\mu}}&
{\bf 1}^{\underline{ab}}{}_{\underline{mn}}
\end{array}
\right)}\label{vielbein}~~~.
\eea
The orthonormal condition \bref{ortho} requires that
``checked variables" are transposed variables plus more many terms,
for example, $\check{B}_{a\mu}=(B_{\mu a} )^T
+\psi_a{}^\alpha B_{\alpha\mu}
-\frac{1}{2}\psi_a{}^\alpha B_{b\alpha}(B_{\mu}{}^b)^T$.
In linearized level
$\check{B}_{\underline{\alpha\mu}}={B}_{\underline{\mu\alpha}}$,
$\check{B}_{\underline{a\mu}}={B}_{\underline{\mu a}}$,
$\check{e}^{\underline{\alpha}}{}_{\underline{\mu}}=
{e}_{\underline{\mu}}{}^{\underline{\alpha}}$, 
$\check{\psi}^{\underline{\alpha m}}
={\psi}^{\underline{m \alpha}}$,
$\check{\omega}^{\underline{ab}}{}^{\underline{\mu}}
={\omega}^{\underline{\mu}}{}^{\underline{ab}}$,
$\check{\omega}^{\underline{ab}}{}^{\underline{m}}
={\omega}^{\underline{m}}{}^{\underline{ab}}$, 
$\check{\omega}^{\underline{ab}}{}_{\underline{\mu}}
={\omega}_{\underline{\mu}}{}^{\underline{ab}}$.

Vielbein fields are classified by canonical dimensions as:
\bea
{\renewcommand{\arraystretch}{1.6}
\begin{array}{l|l}
{\rm type~II~string~field}&{\rm vielbein~field}~({\rm dim.})\\\hline
B~{\rm fields}&B_{\underline{\alpha\mu}}~(-1),~
B_{\underline{a\mu}}~(-\frac{1}{2}),~e_{[\underline{am}]}~(0)\\
{\rm gravity~field}&e_{(\underline{am})},~
e_{\underline{\alpha}}{}^{\underline{\mu}}
~(0)\\
{\rm superconnections}&\omega_{\underline{\alpha}}{}^{\underline{mn}}~(\frac{1}{2})
,~\omega_{\underline{a}}{}^{\underline{mn}}~(1),~
\omega^{\underline{\alpha}}{}^{\underline{mn}}~(\frac{3}{2})\\
{\rm gravitino}&\psi_{\underline{a}}{}^{\underline{\mu}}~(\frac{1}{2})\\
{\rm R-R~field~strength}&F^{{\alpha\mu'}},F^{{\alpha'\mu}}~(1)\\
{\rm covariance~compensator}&r^{\underline{abmn}}~(2)
\end{array}}
\eea
Three kinds of superconnections 
 as well as other fields
 are solved by a set of torsion constraints \bref{2Torsion} resulted from 
the $\kappa$-symmetry.

In linearized level
we have
\bea
E_{\underline{A}}{}^{\underline{M}}&=&\delta_{\underline{A}}^{\underline{M}}
+H_{\underline{A}}{}^{\underline{B}}\delta_{\underline{B}}^{\underline{M}}\nn\\
T_{\underline{ABC}}&=&f_{\underline{ABC}}
+\frac{1}{2}D_{[\underline{A}}H_{\underline{BC})}
+\frac{1}{2}H_{[\underline{A}}{}^{\underline{M}}f_{\underline{BC})\underline{M}}\nn\\
{\cal I}_{\underline{ABCD}}&=&
D_{[\underline{A}}T_{\underline{BCD})}
+\frac{3}{2}T_{[\underline{AB}}{}^{\underline{E}}
f_{\underline{CD})\underline{E}}=0\nn~~~.
\eea
Vielbein fields are solved
  in linearized level;
\bea
{\renewcommand{\arraystretch}{1.6}
\begin{array}{ccl}
{\rm dim}&{\rm torsion~constraint}\Rightarrow&{\rm field}\\
-\frac{1}{2}&T_{{\alpha\beta\gamma}}=0&
B_{a(\alpha}(\gamma^a)_{\beta\gamma)}=D_{(\alpha}B_{\beta\gamma)}\\
&T_{{\alpha\beta\gamma'}}=0
&
B_{a\alpha'}=\frac{1}{8}
D_{\alpha}B_{\beta\gamma'}(\gamma_a)^{\alpha\beta}
+\frac{1}{16}(D_{\gamma'}B_{\alpha\beta})(\gamma_a)^{\alpha\beta}
\\
0&T_{{\alpha\beta c}}=f_{{\alpha\beta c}}&
e_{(\alpha}{}^{\gamma}(\gamma_c)_{\beta)\gamma}
+e_{c}{}^m(\gamma_m)_{\alpha\beta}
=-D_{(\alpha}B_{\beta)c}-D_cB_{\alpha\beta}~\\
&T_{{\alpha\beta c'}}=0&e_{a'}{}^m=\frac{1}{8}
D_\alpha B_{\beta a'}(\gamma^m)^{\alpha\beta}
+\frac{1}{16}D_{a'}B_{\alpha\beta}(\gamma^m)^{\alpha\beta}\\
\frac{1}{2}&T_{{\alpha b c}}=0&
\omega_{{\alpha}}{}_{bc}=-D_\alpha e_{bc}
-D_{[b}B_{c]\alpha}-\psi_{[b}{}^\beta (\gamma_{c]})_{\alpha\beta}
\\
&T_{{\alpha \beta}}{}^{{\gamma}}=0&
\omega_{({\alpha}}{}^{bc}(\gamma_{bc})^\gamma{}_{\beta)}
=-D_{(\alpha} e_{\beta)}{}^{\gamma}+
\psi^{a\gamma} (\gamma_{a})_{\alpha\beta}
 \\ &T_{\alpha bc'}=0&
\psi_{c'}{}^\gamma(\gamma_b)_{\alpha\gamma}=D_{\alpha}e_{bc'}
+D_{[b}B_{c']\alpha}  \\
1& T_{{abc}}=0&~\omega_{abc}=-\frac{1}{3!}D_{[a}e_{bc]}  \\
&T_{a\beta}{}^\gamma=0&
\frac{1}{2}\omega_a{}^{cd}(\gamma_{cd})^\gamma{}_\beta=
-F^{\alpha\gamma}(\gamma_a)_{\beta\alpha}
-D_ae_\beta{}^\gamma+D_\beta\psi_a{}^\gamma\\
\frac{3}{2}&T_{\alpha}{}^{\beta{\gamma}}=0&
\omega^{(\beta| bc}(\gamma_{bc})^{|\gamma)}{}_\alpha=
-4D_\alpha F^{\beta\gamma}-2D^{(\beta}e^{\gamma)}{}_\alpha
~
\end{array}}\label{torsionconstraints}
 \eea
where only  part of relations are listed and 
similar relations for the primed (right) sector are also hold.

Constraints are solved dimension by dimension.
Gauges are chosen so that all of 
$E_{\underline{A}}{}^{\underline{M}}$ 
are fixed in terms of $E_{\underline{D}}{}^{\underline{\Omega}}=
E_{\underline{DD}}=B_{\underline{\alpha\beta}}$
 .  
The dimension $-1$ field, $E_{\underline{DD}}$,
is a prepotential superfield.

Anti-hermiticity of derivative operators is defined by 
$\int e^\phi  \Psi (i{\dd} \chi)=-\int e^\phi   (i{\dd}\Psi) \chi$
with $\dd=\frac{1}{i}\partial+\cdots$,
where the integral measure contains  dilaton $\phi$ which is the only density available.
This leads to a constraint $\tilde{T}_{\underline{A}}$ as
\bea
&&\tilde{T}_{\underline{A}}=
e^{\phi}\overleftarrow{\dd}_{\underline{A}}e^{-\phi}=0\nn\\
&&\tilde{T}_{\underline{\alpha}}=0~\Rightarrow~
\lambda_{\underline{\alpha}}=(D_{\underline{\alpha}}\phi)=
~\frac{i}{4}\omega_{\underline{\beta}}{}^{\underline{bc}}(\gamma_{\underline{bc}})
^{\underline{\beta}}{}_{\underline{\alpha}}
-(D_{\underline{\mu}}e_{\underline{\alpha}}{}^{\underline{\mu}})
-\partial_{\underline{m}}B_{\underline{\alpha}}{}^{\underline{m}}
-B_{\underline{\alpha}}{}^{\underline{m}}(\partial_{\underline{m}}\phi)\nn~~.
\eea
Dilaton $\phi$ and dilatino $\lambda_{\underline{\alpha}}=
D_{\underline{\alpha}}\phi
$
 are introduced through $\tilde{T}_{\underline{\alpha}}=0$.
Equating this constraint  together with 
 torsion constraints  with dimension 1/2
in \bref{torsionconstraints}, 
$T_{\alpha\beta}{}^\beta=
T_{\beta bc}(\gamma^{bc})^\beta{}_\alpha=0$, 
 fixes the trace of 
 $E_{\underline {D\Omega}}=e_{\underline{\alpha}}{}^{\underline{\beta}}$ 
 from $D_{\underline{\alpha}}\phi$ terms.
Analogously $\tilde{T}_{\underline{a}}=0$ constraint together with
the trace of torsion constraints with dimension 1, 
$T_{a\beta}{}^\beta=0$,  
gives an equation for $D_{\underline{\alpha}}D_{\underline{\beta}}\phi$
that fixes $\phi$ as $D_{\underline{\alpha}}D_{\underline{\beta}}
E_{\underline{DD}}
+D_{\underline{\alpha}}D_{\underline{\beta}}V$ 
with some prepotential  $V$
as an homogeneous solution \cite{Pola2014}.

In order to obtain  the usual d-dimensional gravity theory,
dimensional reduction constraints and the section condition 
must be imposed.
Dimensional reduction constraints are written in terms of $\tilde{\dd}_{\underline{M}}$
so that the local  geometry generated by covariant derivatives
is not modified.
The section condition, ${\cal H}_\sigma=0$, is imposed 
in the doubled space with coordinates $(Z^M,Z^{M'})$.
Simple separation of double coordinates into
momenta and $\sigma$ derivative of coordinates
exists.
In this formalism we do not fix the representation of 
R-R field strength as bi-spinor of double Lorentz groups,
instead we leave bi-product of two chiral spinors of
each Lorentz group.
Then dimensional reduction conditions relate two Lorentz groups
and two chiral spinors.
For cases when these conditions mix  chirality under 
T-duality symmetry transformation
IIA and IIB spinors are interchanged.

\par\vskip 6mm

\section{ AdS$^5\times$ S$_5$ superspace}

In this section 
 AdS$^5\times$S$_5$ superspace is explained in a manifestly T-dual formulation, 
 showing the algebra, torsions, curvatures and dimensional reduction
 conditions.
The current algebra for a type IIB superstring in 
AdS$^5\times$S$_5$ background was given in \cite{Hatsuda:2001xf}
in which $\Sigma$ currents  were included while
Lorentz generators were absent  by a constraint $S=0$.
In this section we rederive AdS$^5\times$S$_5$ algebra preserving  
Lorentz generator manifestly 
and  performing current redefinition in such a way that
torsions of the algebra become totally graded antisymmetric.
Then dimensional reduction constraints and section conditions
in our manifestly T-dual formulation are given 
reducing 
 the  AdS$^5\times$S$_5$ algebra in conventional space with
usual coordinates $x^m,\theta_1^\mu, \theta_2^\mu$.

A superstring in AdS$_5\times$S$^5$ 
is described by a coset G/H with 
G=PSU(2,2$\mid$4) and H=SO(5)$\times$SO(4,1).
 Wick rotations change the coset into  G/H with
 G=GL(4$\mid$4) and H=[Sp(4)GL(1)]$^2$.
 The Sp(4) invariant  metric, which is antisymmetric,
 is  denoted by $\hat{\epsilon}$.
The currents and coordinates have
GL(4$\mid$4) index, ${\rm A}=({\rm a},\bar{{\rm a}})$ {\rm a}nd ${\rm a},\bar{{\rm a}}=1,\cdots,4$ ;
\bea
&\displaystyle\frac{\rm G}{\rm H}\ni g_{\rm M}{}^{\rm A}~~~,~~g\to g_G~ g~ g_H~~,~~
(g^{-1})_{\rm A}{}^{\rm M}\partial_\sigma g_{\rm M}{}^{\rm B}=J_{\rm A}{}^{\rm B}~~,~~
\partial_{\rm A}{}^{\rm M} g_{\rm M}{}^{\rm B}=D_{\rm A}{}^{\rm B}~~&\nn
\eea
where $\partial_{\rm A}{}^{\rm M} $ is canonical conjugate of $g_{\rm M}{}^{\rm A}$.
Symmetrization, 
antisymmetrization and traceless antisymmetrization of  indices 
are (ab)=ab+ba, [ab]=ab-ba and $\langle$ab$\rangle$=[ab]-trace part respectively.  
Linear combinations of two GL(4$\mid$4) matrices, $D_{\rm AB}$ and $J_{\rm AB}$,
make 64$\times$2 nondegenerate super-AdS 
currents, $(S,D_\pm,P_\pm,\Omega_\pm,\Sigma,\Delta,\Delta',E,E')$.
This model is rather conventional description so that  
one Lorentz $S$ is involved.
Although there appear  $\Sigma_+$ and $\Sigma_-$ in the algebra,
 only sum $\Sigma=\Sigma_++\Sigma_-$
has nondegenerate commutator with $S$. 
$\Omega_\pm$ is modified from the one in the flat space  denoted by $\omega_\pm$
by the Ramond-Ramond field strengths as 
$\omega_\pm \to \Omega_\pm=\omega_\pm+\frac{1}{2}D_\mp$.
Connection of this model and T-dual formalism is explained 
in the end  of this section.  

The nondegenerate super-AdS currents with GL(4$\mid$4) indies are as follows:
\bea
{\renewcommand{\arraystretch}{1.4}
\begin{array}{lcl}
{\dd}_{\underline M}&\begin{array}{c}
{\rm number~of}\\
{\rm~generators}
\end{array}&{\rm GL}(4{\mid}4)~{\rm currents}\\
\hline
S&20&S_{\rm ab}=\frac{1}{2}{D}_{({\rm ab})},~S_{\bar{{\rm a}}\bar{\rm b}}=
\frac{1}{2}{D}_{(\bar{{\rm a}}\bar{\rm b})}\\
 D_+&16& (D_+){}_{{\rm a}\bar{\rm b}}=E^{1/4}D_{{\rm a}\bar{\rm b}}
 + E^{-1/4}{D}_{\bar{\rm b}{\rm a}}+\frac{1}{2}
 ( E^{1/4}J_{{\rm a}\bar{\rm b}}+ E^{-1/4}{J}_{\bar{\rm b}{\rm a}})\\
D_-&16& (D_-){}_{{\rm a}\bar{\rm b}}=E^{1/4}D_{{\rm a}\bar{\rm b}}- E^{-1/4}{D}_{\bar{\rm b}{\rm a}}
+\frac{1}{2}
 (- E^{1/4}J_{{\rm a}\bar{\rm b}}+ E^{-1/4}J_{\bar{\rm b}{\rm a}})\\
 P_+&10&
  (P_+){}_{{\rm ab}}=\frac{1}{2}
  (D_{\langle {\rm ab}\rangle}+  {J}_{\langle {\rm ab}\rangle}),~
 (P_+){}_{{\rm ab}}= \frac{1}{2}(D_{\langle \bar{{\rm a}}\bar{\rm b}\rangle}+  
J_{\langle \bar{{\rm a}}\bar{\rm b}\rangle})\\
 P_-&10&
 (P_-){}_{{\rm ab}}=\frac{1}{2}(D_{\langle {\rm ab}\rangle}-J_{\langle {\rm ab}\rangle}),~
 (P_{-}){}_{\bar{{\rm a}}\bar{\rm b}}=\frac{1}{2}(D_{\langle \bar{{\rm a}}\bar{\rm b}\rangle}-  
J_{\langle \bar{{\rm a}}\bar{\rm b}\rangle})
\\
{\Omega_+}&16&({\Omega_+})_{{\rm a}\bar{\rm b}}=(\omega_+)_{{\rm a}\bar{\rm b}}+
\frac{1}{2}(D_-)_{{\rm a}\bar{\rm b}}\\
{\Omega_-}&16&({\Omega_-})_{{\rm a}\bar{\rm b}}=(\omega_-)_{{\rm a}\bar{\rm b}}
+\frac{1}{2}(D_+)_{{\rm a}\bar{\rm b}}\nn\\
&&(\omega_+)_{{\rm a}\bar{\rm b}}=E^{1/4}J_{{\rm a}\bar{\rm b}}
 - E^{-1/4}\bar{J}_{\bar{\rm b}{\rm a}}\\
 &&(\omega_-)_{{\rm a}\bar{\rm b}}=-E^{1/4}J_{{\rm a}\bar{\rm b}}
 - E^{-1/4}\bar{J}_{\bar{\rm b}{\rm a}}\\
\Sigma_+&20&(\Sigma_+)_{\rm ab}=\sigma_{\rm ab}+\frac{1}{2}S_{\rm ab}~,~
(\Sigma_+)_{\bar{\rm a}\bar{\rm b}}=\sigma_{\bar{\rm a}\bar{\rm b}}
+\frac{1}{2}S_{\bar{\rm a}\bar{\rm b}}
\\
\Sigma_-&20&(\Sigma_-)_{\rm ab}=\sigma_{\rm ab}-\frac{1}{2}S_{\rm ab}~,~
(\Sigma_-)_{\bar{\rm a}\bar{\rm b}}=\sigma_{\bar{\rm a}\bar{\rm b}}
-\frac{1}{2}S_{\bar{\rm a}\bar{\rm b}}
\\
&&\sigma_{{\rm ab}}=\frac{1}{2}J_{({\rm ab})},~\sigma_{\bar{{\rm a}}\bar{\rm b}}=
 \frac{1}{2}J_{(\bar{{\rm a}}\bar{\rm b})}\\
 \Delta,~\Delta'&2& {\rm Str} D={\rm tr} D_{AdS}-{\rm tr} D_{S},~
 {\rm Tr} D={\rm tr} D_{AdS}+{\rm tr} D_{S}
 \\
E,~E' &2&E={\rm Sdet} g=\displaystyle\frac{{\rm det}g_{AdS}}{{\rm det}g_{S}}
,~E'={\rm Det}g={\rm det}g_{AdS}\cdot {\rm det}g_{S}
\end{array}}\nn\\\label{AdSGL44}
\eea
Algebra of currents in type II 
AdS$^5\times$S$_5$ is given as follows:
\begin{itemize}
  \item {
\textbf{$+~+$ commutators}
\bea
{\renewcommand{\arraystretch}{1.6}
\begin{array}{lcl}
\{(D_+)_{\rm a\bar{\rm b}}(1),(D_+)_{\rm c\bar{d}}(2) \}&=&
-2\left(\hat{\epsilon}_{\rm ac}(P_+)_{\bar{\rm b}\bar{\rm d}}-
\hat{\epsilon}_{\bar{\rm b}\bar{\rm d}}(P_+)_{\rm ac}
+\frac{1}{4}\hat{\epsilon}_{\rm ac}\hat{\epsilon}_{\bar{\rm b}\bar{\rm d}}\Delta
\right)\delta(2-1)\nn\\
\lbrack(D_+)_{\rm a\bar{b}}(1),(P_+)_{\rm cd}(2)]&=&
\hat{\epsilon}_{\rm a\langle c}(\omega_++\frac{1}{2}D_-)_{\rm d\rangle \bar{b}}\delta(2-1)
=
\hat{\epsilon}_{\rm a\langle c}({\Omega_+})_{\rm d\rangle \bar{b}}\delta(2-1)
\nn\\
\{(D_+)_{\rm a\bar{b}}(1),({\Omega_+})_{\rm c\bar{d}}(2)\}&=&
-2\left(\hat{\epsilon}_{\rm ac}({\Sigma_+})_{\rm \bar{b}\bar{d}}+
\hat{\epsilon}_{\rm \bar{b}\bar{d}}({\Sigma_+})_{\rm ac}\right)\delta(2-1)
+2\hat{\epsilon}_{\rm ac}\hat{\epsilon}_{\rm \bar{b}\bar{d}}\partial_\sigma\delta(2-1)
\nn\\
\lbrack(P_+)_{\rm a{b}}(1),(P_+)_{\rm cd}(2)]&=&
\hat{\epsilon}_{\rm \langle a\mid \langle c}({\Sigma_+})_{\rm d\rangle\mid
 {b}\rangle}\delta(2-1)
+\hat{\epsilon}_{\rm a\langle c}\hat{\epsilon}_{\rm {d}\rangle {b}}
 \partial_\sigma\delta(2-1)
\nn\\
\lbrack(P_+)_{\rm a{b}}(1),({\Omega_+})_{\rm c\bar{d}}(2)]&=&
\frac{1}{2}\hat{\epsilon}_{\rm c \langle a}({\Omega_-}-D_+)
_{\rm b\rangle\mid \bar{d}}\delta(2-1)\nn\\
\{({\Omega_+})_{\rm a\bar{b}}(1),({\Omega_+})_{\rm c\bar{d}}(2)\}&=&
\left(\hat{\epsilon}_{\rm ac}(P_+-\frac{1}{2}P_-)_{\rm \bar{b}\bar{d}}-
\hat{\epsilon}_{\rm \bar{b}\bar{d}}(P_+-\frac{1}{2}P_-)_{\rm ac}
+\frac{1}{8}\hat{\epsilon}_{\rm ac}\hat{\epsilon}_{\bar{\rm b}\bar{\rm d}}\Delta
\right)\delta(2-1)
\end{array}}
\label{plpl}
\eea
This is summarized in the table of indices of structure constant
$f_{AB}{}^C=
\begin{array}{|c|c|}
\hline
&B\\\hline
A&C\\
\hline
\end{array}
$
\bea
{\renewcommand{\arraystretch}{1.4}
\begin{array}{|c|c|c|c|}
\hline
&~~~~~D_+~~~~~&~~~~~P_+~~~~~&~~~~~{\Omega_+}~~~~~\\
\hline
D_+&P_+&{\Omega_+}&\Sigma_+,\partial_\sigma\delta\\
\hline
P_+&\Omega_+&\Sigma_+,\partial_\sigma\delta&D_+,{\Omega_-}\\
\hline
{\Omega_+}&\Sigma_+,\partial_\sigma\delta
&D_+,{\Omega_-}&P_+,P_-\\
\hline
\end{array}}\label{plapla}
\eea

}
  \item {
\textbf{$-~- $ commutators}
\bea
{\renewcommand{\arraystretch}{1.6}
\begin{array}{lcl}
 \{(D_-)_{\rm a\bar{b}}(1),(D_-)_{\rm c\bar{d}}(2)\}&=&
2\left(\hat{\epsilon}_{\rm ac}(P_-)_{\rm \bar{b}\bar{d}}-
\hat{\epsilon}_{\bar{b}\bar{d}}(P_-)_{\rm ac}
+\frac{1}{4}\hat{\epsilon}_{\rm ac}\hat{\epsilon}_{\bar{\rm b}\bar{\rm d}}\Delta
\right)\delta(2-1)\nn\\
\lbrack(D_-)_{\rm a\bar{b}}(1),(P_-)_{\rm cd}(2)]&=&
\hat{\epsilon}_{\rm a\langle c}(\omega_-
+\frac{1}{2}D_+)_{\rm d\rangle \bar{b}}\delta(2-1)=
\hat{\epsilon}_{\rm a\langle c}({\Omega_-})_{\rm d\rangle \bar{b}}\delta(2-1)
\nn\\
\{(D_-)_{\rm a\bar{b}}(1),({\Omega_-})_{\rm c\bar{d}}(2)\}&=&
2\left(\hat{\epsilon}_{\rm ac}({\Sigma_-})_{\rm \bar{b}\bar{d}}+
\hat{\epsilon}_{\rm \bar{b}\bar{d}}({\Sigma_-})_{\rm ac}\right)\delta(2-1)
+2\hat{\epsilon}_{\rm ac}\hat{\epsilon}_{\rm \bar{b}\bar{d}}\partial_\sigma\delta(2-1)
\nn\\
\lbrack(P_-)_{\rm a{b}}(1),(P_-)_{\rm cd}(2)]&=&
-\hat{\epsilon}_{\rm \langle a\mid \langle c}({\Sigma_-})_{\rm d\rangle\mid
 {b}\rangle}\delta(2-1)
 -\hat{\epsilon}_{\rm a\langle c}\hat{\epsilon}_{\rm {d}\rangle {b}}
 \partial_\sigma\delta(2-1)
\nn\\
\lbrack(P_-)_{\rm a{b}}(1),({\Omega_-})_{\rm c\bar{d}}(2)]&=&
\frac{1}{2}\hat{\epsilon}_{\rm c \langle a}({\Omega_+}-D_-)
_{\rm b\rangle\mid \bar{d}}\delta(2-1)\nn\\
\{({\Omega_-})_{\rm a\bar{b}}(1),({\Omega_-})_{\rm c\bar{d}}(2)\}&=&
\left(-\hat{\epsilon}_{\rm ac}(P_--\frac{1}{2}P_+)_{\rm \bar{b}\bar{d}}+
\hat{\epsilon}_{\rm \bar{b}\bar{d}}(P_--\frac{1}{2}P_+)_{\rm ac}
-\frac{1}{8}\hat{\epsilon}_{\rm ac}\hat{\epsilon}_{\bar{\rm b}\bar{\rm d}}\Delta
\right)
\delta(2-1)\nn\end{array}
\label{mimi}}
\eea
The structure constant among 
the $-$ currents is summarized as
\bea
{\renewcommand{\arraystretch}{1.4}
\begin{array}{|c|c|c|c|}
\hline
&~~~~~D_-~~~~~&~~~~~P_-~~~~~&~~~~~{\Omega_-}~~~~~\\
\hline
D_-&P_-&{\Omega_-}&\Sigma_-,\partial_\sigma\delta\\
\hline
P_-&\Omega_-&\Sigma_-,\partial_\sigma\delta&D_-,{\Omega_+}\\
\hline
{\Omega_-}&\Sigma_-,\partial_\sigma\delta
&D_-,{\Omega_+}&P_+,P_-\\
\hline
\end{array}}\label{minmin}
\eea
}
\item{
\textbf{$+~-$ commutators}
\bea
{\renewcommand{\arraystretch}{1.6}
\begin{array}{lcl}
\{(D_+)_{\rm a\bar{b}}(1),(D_-)_{\rm c\bar{d}}(2)\}&=&
-2\left(\hat{\epsilon}_{\rm ac}S_{\rm \bar{b}\bar{d}}-
\hat{\epsilon}_{\rm \bar{b}\bar{d}}S_{\rm ac}\right)\delta(2-1)\nn\\
\lbrack(D_+)_{\rm a\bar{b}}(1),(P_-)_{\rm cd}(2)]&=&
\hat{\epsilon}_{\rm a\langle c}(D_-)_{\rm d\rangle \bar{b}}
\delta(2-1)
\nn\\
\{(D_+)_{\rm a\bar{b}}(1),({\Omega_-})_{\rm c\bar{d}}(2)\}&=&
-\left(\hat{\epsilon}_{\rm ac}(P_-)_{\rm \bar{b}\bar{d}}-
\hat{\epsilon}_{\rm \bar{b}\bar{d}}(P_-)_{\rm ac}
+\frac{1}{4}\hat{\epsilon}_{\rm ac}\hat{\epsilon}_{\bar{\rm b}\bar{\rm d}}\Delta
\right)\delta(2-1)
\nn\\
\lbrack(P_+)_{\rm a{b}}(1),(P_-)_{\rm cd}(2)]&=&
\frac{1}{2}\hat{\epsilon}_{\rm \langle a\mid \langle c}
S_{\rm d\rangle\mid {b}\rangle}\delta(2-1)\nn\\
\lbrack(P_+)_{\rm a{b}}(1),({\Omega_-})_{\rm c\bar{d}}(2)]&=&
-\frac{1}{2}\hat{\epsilon}_{\rm c \langle a}(D_-)
_{\rm b\rangle\mid \bar{d}}\delta(2-1)\nn\\
\left\{({\Omega_+})_{\rm a\bar{b}}(1),({\Omega_-})_{\rm c\bar{d}}(2)\right\}&=&
\frac{1}{2}\left(\hat{\epsilon}_{\rm ac}S_{\rm \bar{b}\bar{d}}-
\hat{\epsilon}_{\rm \bar{b}\bar{d}}S_{\rm ac}\right)\delta(2-1)\label{plmi}\\
\lbrack(P_+)_{\rm a{b}}(1),(D_-)_{\rm c\bar{d}}(2)]&=&
-\frac{1}{2}\hat{\epsilon}_{\rm c\langle a}
(D_+)_{\rm b\rangle \bar{d}}\delta(2-1)\nn\\
\lbrack({\Omega_+})_{\rm a\bar{b}}(1),(P_-)_{\rm cd}(2)]&=&
\frac{1}{2}\hat{\epsilon}_{\rm a\langle c}
(D_+)_{\rm d\rangle \bar{b}}\delta(2-1)
\end{array}}
\eea
The structure constant between the $+$ current and $-$ currents is summarized as
\bea
{\renewcommand{\arraystretch}{1.4}
\begin{array}{|c|c|c|c|}
\hline
&~~~~~D_-~~~~~&~~~~~P_-~~~~~&~~~~~{\Omega_-}~~~~~\\
\hline
D_+&S&D_-&P_-\\
\hline
P_+&D_+&S&D_-\\\hline
{\Omega_+}&P_+&D_+&S\\
\hline
\end{array}}\label{plami}
\eea
}
\end{itemize}
Non-vanishing torsions are 
\bea
{\renewcommand{\arraystretch}{1.6}
\begin{array}{lcl}
{\rm dim}~0&:&
T_{DDP}=f_{DDP}~~,~~T_{PPS}=f_{PPS}~~,~~
T_{D\Omega S }=f_{D\Omega S}\nn\\
{\rm dim}~1&:&T_{P\Omega D'},~T_{DD'}{}^S=R_{DD'}{}^S\nn\\
{\rm dim}~2&:&
T_{\Omega\Omega P},~T_{\Omega \Omega P'}~,
~T_{PP'}{}^S=R_{PP'}{}^S \nn\\
{\rm dim}~3&:&T_{\Omega \Omega'}{}^S=R_{\Omega \Omega'}{}^S
\end{array}}\nn
\eea
These torsions are consistent with torsion constraints to
\bref{Torsion1}, 
\bref{2Torsion} and
\bref{TSTS}.

Let us examine the current algebra 
 AdS$_5\times$S$^5$ by translating GL(4${\mid}$4) indices into 
 spacetime indices, then write down  
 Bianchi identities
\bea
{\rm dim.}~2~\lbrack D_\alpha,D_{\beta'},P_a]=0~&\Rightarrow& ~
R_{\alpha\beta'}{}^{ab}=T_{\beta'}{}^{\beta[a} (\gamma^{b]})_{\alpha\beta}
\nn\\
\lbrack D_\alpha,D_\beta,P_{c'}]=0~&\Rightarrow&~
R_{aa'}{}^{bc}(\gamma^a)_{\alpha\beta}=
T_{a'(\alpha|}{}^{\gamma'}R_{\gamma'|\beta)}{}^{bc}\nn\\
{\rm dim.}~3~
\lbrack D_\alpha,P_{b},\Omega^{\gamma'}]=0~&\Rightarrow&~
R^{\beta\gamma';cd }(\gamma_b)_{\alpha\beta}
=T^{c'\gamma'}{}_{\alpha}R_{c'b}{}^{cd}
+T_{b}{}^{\gamma'\gamma}R_{\gamma'\alpha}{}^{cd}\nn
\eea
Torsions with dimension 1 are written in terms of
 R-R field strength $F^{\alpha\beta'}$ with
 dimension 1 as
\bea
T_{a\alpha'}{}^{\beta}=(\gamma_a)_{\alpha'\beta'}F^{\beta\beta'}~,~
T_{a'\alpha}{}^{\beta'}=(\gamma_a)_{\alpha\beta}F^{\alpha\beta'}~.\nn
\eea
The R-R field strength of the AdS$_5\times$S$^5$ space 
is given by
$F^{\alpha\beta'}=\frac{1}{r_{\rm AdS}}(\Gamma_5)^{\alpha\beta'}$
with the  AdS radius, $r_{\rm AdS}$.
Prepotential $B_{\alpha\beta'}$ has terms such as 
$\frac{1}{g}F$,
 $F\theta{}^2x$  and  $F\theta{}^4$ where $g$ is string coupling.

This description is in conventional space with coordinates
$(x^m,~\theta^\mu,~\theta^{\mu'})$,
so we must reduce dimensions from the doubled space.
Dimensional reduction is imposed using symmetry generators $\tilde{\dd}_M$ 
without modifying local  algebras of covariant derivatives.
We impose set of dimensional reduction constraints
\bea
&\tilde{P}-\tilde{P'}=
\tilde{S}-(-\tilde{S'})=0~,~
\tilde{\Omega}=\tilde{\Omega'}=
\tilde{\Sigma}=\tilde{\Sigma'}=0~~~~,&
\eea
which reduce into one set of nondegenerate superspace.
Physical translation generator is
$\tilde{P}_m+\tilde{P}'{}_{m'} \to \tilde{P}_m$,
while usual coordinate is $x^m+x'{}^m \to x^m$ .
Physical Lorentz generator is
$\tilde{S}_{mn}+(-\tilde{S}'{}_{m'n'}) \to \tilde{S}_{mn}$.
Lorentz coordinates are suppressed by the gauge fixing condition 
$u^{mn}=u^{m'n'}=0$ corresponding to first class constraints
$S=S'=0$. 
Type IIB fermionic coordinates are chosen as follows,
resulting the  conventional Lorentz symmetry generator;
\bea
&
\lbrack \tilde{S}_{mn},\theta^\mu]=\frac{i}{2}(\gamma_{mn})^{\mu}{}_{\nu}\theta^\nu~,~
\lbrack \tilde{S}_{mn},\theta^{\mu'}]=\frac{i}{2}(\gamma_{mn})^{\mu}{}_{\nu}\theta{}^{\nu'}~,~
{\Gamma}_{11}\theta=\theta~,~
{\Gamma}_{11}\theta'=\theta'&\nn\\
&\tilde{S}_{mn}=ix_{[m}\partial_{n]}-i
\theta^\mu\partial_\nu(\gamma_{mn})^\nu{}_\mu
-i\theta^{\mu'}\partial{}_{\nu'}(\gamma_{mn})^\nu{}_\mu~~~.&\nn
\eea

\par
\vskip 6mm
\section{Conclusions}

We present a superspace formulation of type II superstring
background with manifest T-duality symmetry by
following the procedure given in \cite{Polacek:2013nla}.
The nondegenerate super-Poincar\'{e} algebra is given 
in \bref{SDPomesig} to define the superspace.
The $\kappa$-symmetric Virasoro constraints, 
${\cal ABCD}$ constraints, in the nondegenerate super-Poincar\'{e} 
currents are also given in
\bref{ABCD} and \bref{ABCDalg}.
Then we double whole set of nondegenerate super-Poincar\'{e}
algebra to construct type II theory
with manifest T-duality.

The vielbein superfield $E_{\underline{AM}}$ is 
introduced in \bref{vielbein} which contains all fields.
Torsion constraints are obtained from  $\kappa$-symmetric Virasoro constraints in \bref{2Torsion}.
These torsion constraints are solved, and 
all superconnections and vielbein fields are 
written in terms of a prepotential $E_{\underline{\alpha\beta}}$.

This formulation is a double superspace field theory 
with two sets of coordinates in \bref{SDPWZ}.
Gauge fixing for coset constraints $S=0$ allows
to gauge away $u$ Lorentz coordinates.
Dimensional reduction constraints $\tilde{\Omega}=\partial_\varphi+...=0$ 
and  $\tilde{\Sigma}=\partial_v+...=0$ 
allow  to remove $\varphi$ and $v$.
The space has the  O(d,d) T-duality 
symmetry as well as supersymmetry.
Further dimensional reductions of Lorenz and translation generators 
and gauge fixing of the section condition
are imposed to obtain the usual coordinate space.

We leave  interesting aspects
such as R-R gauge fields, D-branes, 
 interchanging IIA and IIB and M-theory,
 for future problem. 
The manifestly T-dual formulation in other gauge fixing 
may allow to explore further,
such as exotic branes and low energy effective theory of F-theory.

\par\vskip 6mm

\section*{Acknowledgements}
We thank  to Chris Hull and 
 Martin Pol\'{a}{\v{c}}ek for valuable discussions. 
M.H. would like to thank the Simons Center for Geometry and Physics for
hospitality during the 
2013 Summer Simons workshop in Mathematics and Physics
where this work has been developed. 
 The work of M.H. is supported  by Grant-in-Aid for Scientific Research (C) No. 24540284 from The Ministry of Education, Culture, Sports, Science and Technology of Japan,
and the work of W.S. is
 supported in part by National Science Foundation Grant No. PHY-1316617.



\vskip 6mm

\begin{thebibliography}{99}

  
   \bibitem{Siegel:1993xq}
  W.~Siegel,
  ``Two vierbein formalism for string inspired axionic gravity,''
  Phys.\ Rev.\ D {\bf 47} (1993) 5453
  [hep-th/9302036];
  ``Superspace duality in low-energy
superstrings,'' Phys.\ Rev.\ D {\bf 48} (1993) 2826 [hep-th/9305073];
  ``Manifest T-duality in low-energy superstrings,''
  [arXiv:hep-th/9308133].

  \bibitem{Hitchin:2004ut}
  N.~Hitchin,
  ``{ Generalized Calabi-Yau manifolds},''
  Quart.\ J.\ Math.\ Oxford Ser.\  {\bf 54} (2003) 281
  [arXiv:math/0209099];\\
  M.~Gualtieri,
  ``{ Generalized complex geometry},''
  math/0401221 [math-dg].
    
  \bibitem{Hull:2004in}
  C.~M.~Hull,
  ``A Geometry for non-geometric string backgrounds,''
  JHEP {\bf 0510} (2005) 065
  [hep-th/0406102];
``Doubled Geometry and T-Folds,''
  JHEP {\bf 0707} (2007) 080
  [hep-th/0605149];
  ``{ Generalised Geometry for M-theory},''
  JHEP {\bf 0707} (2007) 079
  [hep-th/0701203].
  
\bibitem{Grana:2005jc}
  M.~Grana,
  ``Flux compactifications in string theory: A Comprehensive review,''
  Phys.\ Rept.\  {\bf 423} (2006) 91
  [hep-th/0509003].
  
    \bibitem{Hohm:2013bwa}
  O.~Hohm, D.~Lust and B.~Zwiebach,
  ``The Spacetime of Double Field Theory: Review, Remarks, and Outlook,''
  Fortsch.\ Phys.\  {\bf 61} (2013) 926
  [arXiv:1309.2977 [hep-th]].
  

\bibitem{Berman:2014jba}
  D.~S.~Berman, M.~Cederwall and M.~J.~Perry,
  ``Global aspects of double geometry,''
  arXiv:1401.1311 [hep-th].
  

  
    \bibitem{Hassan:1999bv}
  S.~F.~Hassan,
    ``SO(d,d) transformations of Ramond-Ramond fields and space-time spinors,''
  Nucl.\ Phys.\ B {\bf 583} (2000) 431
  [hep-th/9912236].

\bibitem{Fukuma:1999jt}
  M.~Fukuma, T.~Oota and H.~Tanaka,
  ``Comments on T dualities of Ramond-Ramond potentials on tori,''
  Prog.\ Theor.\ Phys.\  {\bf 103} (2000) 425
  [hep-th/9907132].
  
    \bibitem{Hohm:2011zr}
  O.~Hohm, S.~K.~Kwak and B.~Zwiebach,
  ``Unification of Type II Strings and T-duality,''
  Phys.\ Rev.\ Lett.\  {\bf 107} (2011) 171603
  [arXiv:1106.5452 [hep-th]].
  
    \bibitem{Jeon:2012kd}
  I.~Jeon, K.~Lee and J.~-H.~Park,
   ``Ramond-Ramond Cohomology and O(D,D) T-duality,'' JHEP {\bf 1209} (2012) 079
  [arXiv:1206.3478 [hep-th]];\\
 I.~Jeon, K.~Lee, J.~-H.~Park and Y.~Suh,
  ``Stringy Unification of Type IIA and IIB Supergravities under N=2 D=10 Supersymmetric Double Field Theory,''
  Phys.\ Lett.\ B {\bf 723} (2013) 245
  [arXiv:1210.5078 [hep-th]].

 


  
\bibitem{Siegel:85}
 W. Siegel, 
 ``Covariant approach to superstrings, h{\it in}  Symposium on
anomalies, geometry, topology, Chicago, March 27-30, 1985,
eds. W.A. Bardeen and A.R. White (World Scientific, Singapore, 1985)
348;\\
Covariant superstrings, {\it in}  Unified string theories, Santa Barbara,
July 29 - August 16, 1985, eds. M. Green and D. Gross
(World Scientific, Singapore, 1985) 559.

\bibitem{Siegel:1985xj}
  W.~Siegel,
  ``Classical Superstring Mechanics,''
  Nucl.\ Phys.\ B {\bf 263} (1986) 93.

\bibitem{Hatsuda:2001xf}
  M.~Hatsuda and K.~Kamimura,
  ``Classical AdS superstring mechanics,''
  Nucl.\ Phys.\ B {\bf 611} (2001) 77
  [hep-th/0106202];\\
  M.~Hatsuda,
  ``Sugawara form for AdS superstring,''
  Nucl.\ Phys.\ B {\bf 730} (2005) 364
  [hep-th/0507047].
  
  \bibitem{Bonanos:2009wy}
  S.~Bonanos, J.~Gomis, K.~Kamimura and J.~Lukierski,
  ``Maxwell Superalgebra and Superparticle in Constant Gauge Badkgrounds,''
  Phys.\ Rev.\ Lett.\  {\bf 104}, 090401 (2010)
  [arXiv:0911.5072 [hep-th]].
  
  
\bibitem{Siegel:2011sy}
  W.~Siegel,
  ``New superspaces/algebras for superparticles/strings,''
  arXiv:1106.1585 [hep-th].
    
   \bibitem{Polacek:2013nla}
  M.~Pol\'{a}{\v{c}}ek and W.~Siegel,
  ``Natural curvature for manifest T-duality,''
  JHEP {\bf 1401} (2014) 026
  [arXiv:1308.6350 [hep-th]].
    
 
\bibitem{Howe:1981gz}
  P.~S.~Howe,
  ``Supergravity in Superspace,''
  Nucl.\ Phys.\ B {\bf 199} (1982) 309.


\bibitem{Howe:1983sra}
  P.~S.~Howe and P.~C.~West,
  ``The Complete N=2, D=10 Supergravity,''
  Nucl.\ Phys.\ B {\bf 238} (1984) 181.
  
\bibitem{Berkovits:2001ue}
  N.~Berkovits and P.~S.~Howe,
  ``Ten-dimensional supergravity constraints from the pure spinor formalism for the superstring,''
  Nucl.\ Phys.\ B {\bf 635} (2002) 75
  [hep-th/0112160].
     
            

 \bibitem{Hatsuda:2012uk}
  M.~Hatsuda and T.~Kimura,
  ``Canonical approach to Courant brackets for D-branes,''
  JHEP {\bf 1206} (2012) 034
  [arXiv:1203.5499 [hep-th]].
  
  \bibitem{Hatsuda:2012vm}
  M.~Hatsuda and K.~Kamimura,
  ``SL(5) duality from canonical M2-brane,''
  JHEP {\bf 1211} (2012) 001
  [arXiv:1208.1232 [hep-th]];
  ``M5 algebra and SO(5,5) duality,''
  JHEP {\bf 1306} (2013) 095
  [arXiv:1305.2258 [hep-th]].


\bibitem{Witten:1985nt}
  E.~Witten,
  ``Twistor - Like Transform in Ten-Dimensions,''
  Nucl.\ Phys.\ B {\bf 266} (1986) 245.
  

\bibitem{Shapiro:1986xp}
  J.~A.~Shapiro and C.~C.~Taylor,
  ``Supergravity Torsion Constraints From the 10-$D$ Superparticle,''
  Phys.\ Lett.\ B {\bf 181} (1986) 67;
   ``Superspace Supergravity From the Superstring,''
  Phys.\ Lett.\ B {\bf 186} (1987) 69.

   
  \bibitem{Pola2014}
  M.~Pol\'{a}{\v{c}}ek and W.~Siegel,
  ``T-duality off shell in 3D Type II superspace,''
  arXiv:1403.6904 [hep-th].
  
\end{thebibliography}
\end{document}